\documentclass[a4paper,fleqn,usenatbib]{mnras}

 \usepackage{newtxtext,newtxmath}

\usepackage[T1]{fontenc}
\usepackage{ae,aecompl}

\usepackage{graphicx}	
\usepackage{amsmath}	
\usepackage{amssymb}	

\usepackage{bm}
\newcommand\fx{f_{\rm x}}
\newcommand\px{P_{\rm x}}
\newcommand\ax{A_{\rm x}}
\newcommand\fo{f_{\rm 1O}}
\newcommand\po{P_{\rm 1O}}
\renewcommand\ao{A_{\rm 1O}}

\title[Double-mode radial-non-radial RR Lyrae stars]{Double-mode radial--non-radial RR Lyrae stars. OGLE-IV photometry of two high cadence fields in the Galactic bulge.}
 \author[H. Netzel, R. Smolec \& P. Moskalik]
 {H. Netzel$^{1}$\thanks{E-mail: henia@netzel.pl},
 R. Smolec$^{2}$\thanks{E-mail: smolec@camk.edu.pl} and
 P. Moskalik$^{2}$\\
 $^{1}$Instytut Astronomiczny, Uniwersytet Wroc\l{}awski, ul. Kopernika 11, 51-622 Wroc\l{}aw, Poland\\
 $^{2}$Nicolaus Copernicus Astronomical Centre, Polish Academy of Sciences, Bartycka 18, 00-716 Warszawa, Poland\\
 }

\date{Accepted . Received ; in original form }

\pubyear{2015}

\begin{document}
\label{firstpage}
\pagerange{\pageref{firstpage}--\pageref{lastpage}} 
\maketitle

\begin{abstract}
We analyse the OGLE-IV photometry of the first overtone and double-mode RR~Lyrae stars (RRc/RRd) in the two fields towards the Galactic bulge observed with high cadence. In 27 per cent of RRc stars we find additional non-radial mode, with characteristic period ratio, $\px/\po\in(0.6, 0.64)$. It strongly corroborates the conclusion arising from the analysis of space photometry of RRc stars, that this form of pulsation must be common. In the Petersen diagram the stars form three sequences. In 20 stars we find two or three close secondary modes simultaneously. The additional modes are clearly non-stationary. Their amplitude and/or phase vary in time. As a result, the patterns observed in the frequency spectra of these stars may be very complex. In some stars the additional modes split into doublets, triplets or appear as a more complex bands of increased power. Subharmonics of additional modes are detected in 20 per cent of stars. They also display a complex structure.

Including our previous study of the OGLE-III Galactic bulge data, we have discovered 260 RRc and 2 RRd stars with the additional non-radial mode, which is the largest sample of these stars so far. The additional mode is also detected in two Blazhko RRc stars, which shows that the modulation and additional non-radial mode are not exclusive.

\end{abstract}




\section{Introduction}\label{sec:intro}
RR~Lyrae stars are considered to be textbook examples of simple,
classical, radially pulsating variables. Indeed, majority of these
stars pulsate in the fundamental mode (F mode, RRab stars) or in the
first radial overtone (1O mode, RRc). A less frequent is
simultaneous pulsation in these two radial modes (F+1O, RRd stars).
Till very recently, a still mysterious quasi-periodic modulation of
the pulsation amplitude and/or phase -- the Blazhko effect
\citep[for a review see e.g.][]{szabo14} -- was the only flaw in
this simple picture.

With the advent of space based photometry and overwhelming amount of
data gathered by ground-based photometric surveys, in particular by
the Optical Gravitational Lensing Experiment
\citep[OGLE,][]{ogleIII,ogleiv_tech}, as well as by the dedicated ground
based campaigns \citep[e.g.][]{jurcsik_KBS}, the simple picture
outlined above does not hold. Several stars pulsating simultaneously
in the fundamental and in the second overtone were detected
\citep[for a review see][]{pam13}. Many interesting discoveries in
Blazhko variables were made, including discovery of period doubling
in modulated RRab stars \citep{kol10,szabo10} or discovery of the
Blazhko effect in RRd variables \citep{ogleiv,rs15a,jurcsik_BLRRd}.

The most intriguing discovery however, seems to be the detection of
non-radial modes in RRc stars, likely a common feature of these
stars. The period of additional mode, $P_{\rm x}$, is shorter than
first overtone period; the period ratios, $P_{\rm x}/P_{\rm 1O}$,
fall in a narrow range between $0.60$ and $0.64$. The first
detection of non-radial mode of this type was made in RRd star
AQ~Leo observed with the {\it MOST} satellite \citep{aqleo}. Then,
the same additional mode was also detected in 6 RRc stars in the
globular cluster omega Centauri \citep{om09}. Other RRc stars with
similar period ratios were discovered in the OGLE LMC data
\citep{ogle_rr_lmc} and in the SDSS data \citep{sdss}. A big
surprise was the analysis of 4 RRc stars observed with {\it Kepler}
space telescope \citep{pamsm15}. In all these stars additional mode
with characteristic period ratio in a range $\sim(0.61,\, 0.63)$ was
detected. Analysis of the {\it CoRoT} photometry \citep{szabo_corot}
and {\it K2}\footnote{{\it K2} is the continuation of {\it Kepler}
space telescope mission, with observations carried along ecliptic
plane, after the second reaction wheel failure.} observations
\citep{molnar} leave no doubt: in 13 out of 14 RRc stars observed
from space the additional mode is detected\footnote{The only RRc
star observed from space in which additional mode is not detected
shows the Blazhko effect. Still, only 8.9\thinspace d of
observations are available for this star and hence the presence of
additional $0.61$ mode cannot be excluded, specially taking into
account its variable nature, see Section~\ref{ssec:seasonal}.}
 -- the phenomenon must be common in RRc stars.

Possible existence of a new group of double-periodic pulsators with
period ratio close to $0.61$ was postulated already by \cite{om09}
and fully confirmed with the analysis of the {\it Kepler} photometry
\citep{pam+13,pamsm15}. In all these stars first overtone pulsation is
dominant. The additional mode may also occur in RRd variables.
Sub-harmonics of the additional mode are detected in the majority of
stars observed from space and seem characteristic for this group. In
the following we will refer to these stars as $0.61$-stars. The
period ratio, cannot correspond to two radial modes, as model
computations clearly show -- it is in between period ratios expected
for 1O+3O and 1O+4O pulsators \citep{pamsm15}.

The additional mode in $0.61$-stars is of low amplitude, in the mmag
range, typically, 1 to 4 per cent of the first overtone amplitude.
The low amplitude makes the detection of additional mode difficult
from the ground. Nevertheless, in our analysis of the OGLE-III
photometry of the Galactic bulge \citep{netzel}, we have found 147
RRc and RRd stars with the additional mode (3 per cent of OGLE-III
RRc sample), increasing the number of known stars of this type by
factor 6 and allowing first statistical analyses. In the Petersen
diagram majority of these stars form a tight sequence with period
ratios clustering around $0.613$. A signature of a second sequence
with slightly larger period ratio ($0.63$) was also detected.

Here we report the analysis of the top-quality OGLE-IV data for the Galactic bulge RRc stars. A detection of 131 stars, which is 27 per cent of the analysed sample, together with recent results of  \cite{jurcsik_M3}, who report the detection of eighteen 0.61 stars in M3 ($\sim$ 38 per cent of their sample), corroborate the results of space mission -- the phenomenon must be common among RRc stars. Our new results allow much more detailed analysis of the group. In several stars we detect very rich structures in the frequency spectrum. At the frequency range characteristic for the additional mode we observe one, two or even three well separated peaks; corresponding period ratios are in a range $\sim(0.61-0.63)$. These peaks give rise to three sequences in the Petersen diagram -- in addition to two sequences we detected before, a third, although scarcely populated, clearly appears in between. Rich and complex structures are detected at frequency range around sub-harmonic of the additional mode(s).

\section{Data and analysis}\label{sec:analysis}
\begin{figure}
\centering
\resizebox{\hsize}{!}{\includegraphics{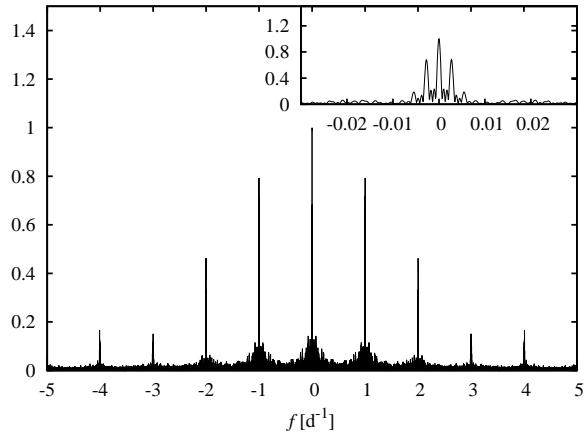}}
\caption{Example of spectral window for OGLE-IV data.}
\label{fig.win}
\end{figure}

In the analysis we used data from the fourth phase of the OGLE project \citep{ogleiv,ogleiv_tech}. In publicly available OGLE-IV data\footnote{\textsf{ftp://ftp.astrouw.edu.pl/ogle/ogle4\/}} there are four observational seasons covering $1334$\thinspace days for most stars. OGLE-IV collection of variable stars contains $38\,257$ RR~Lyrae stars from the Galactic bulge, including $10\,825$ RRc stars and $174$ RRd stars. For the analysis we selected stars pulsating in the first overtone (RRc and RRd) placed in the most frequently observed fields (501 and 505 in OGLE-IV). This choice is motivated by our goal, search for low amplitude signals, which requires possibly the lowest noise level. This is met by stars with most numerous observations. There are more than $8\,000$ data points for each star. Selected sample consist of 485 RRc and 4 RRd stars. We used only $I$-band data, as they are much more numerous than $V$-band data. Spectral window for OGLE-IV data is presented in Fig.~\ref{fig.win}. It is typical for the OGLE photometry of the Galactic bulge. Strong 1-day and 1-year aliases are present.

All data were analysed manually. In the analysis we used standard successive prewhitening method. Significant frequencies were found with discrete Fourier transform and fitted to the data in the form:
\begin{equation} \label{eq.fit}
 m(t)=m_0+\sum_{k=1}^N{A_k\sin(2\pi f_k t + \phi_k)}\,,
\end{equation}
where $f_k$ are frequencies, $A_k$ and $\phi_k$ are amplitudes and phases. We considered only frequencies with signal-to-noise ratio $S/N \geqslant 4$. Then we performed detrending and data clipping with $4\sigma$ criterion, where $\sigma$ is the dispersion of the fit.

Slow trends are present in the data of many stars. In the frequency spectrum they produce signal at low frequencies which gives rise to daily aliases at around integer frequency values. We modelled these trends with long-period sine function ($50\,000$\thinspace d) or we removed them from the data with low-order polynomial.

In many stars, after prewhitening with the frequency of the first overtone, $f_{\rm 1O}$, signals in the vicinity of $f_{\rm 1O}$ remain. They form either doublets of equidistant triplets and multiplets at the main frequency and its harmonics. These are manifestations of the Blazhko effect in the frequency spectrum. We fitted these signals in the form $kf_{\rm 1O} \pm n\Delta f$, where $\Delta f$ is a separation between main frequency and the side peaks.

In some stars signals at the location of $f_{\rm 1O}$ or its harmonics remained despite prewhitening. It is caused by the change of amplitude and/or phase of the first overtone on time-scale longer than data length. Non-stationary signals result in higher noise level, give raise to daily aliases, and so hamper the search for additional low amplitude signals. In order to remove such signals we used time-dependent prewhitening method proposed by \cite{pamsm15}. Its application to OGLE data is described in more detail in \cite{netzel}. In cases of stars for which OGLE-III data \citep{ogle_rr_blg} are available we merged the data in order to investigate long-term changes of the first overtone. Irregular phase variations are frequent, but in some cases analysis of the merged data revealed long-period Blazhko effect.

For all stars for which we found additional periodicity of interest, $f_{X}$, we conducted seasonal analysis. Analysis, as described above, was performed on each of the four observing seasons separately in order to investigate time variation of the additional periodicity (Section~\ref{ssec:seasonal}).

In this paper our attention is focused only on 0.61-stars. In the same sample of stars we have detected yet another intriguing group of double-periodic radial--non-radial variables, with period of the additional mode longer than first overtone period. This result was reported in \cite{netzel68}. Detailed analysis of the Blazhko effect in RRc stars is in preparation.

\section{Results}\label{sec:results}
\subsection{Overview}

We have found 131 RRc stars with the additional non-radial mode of
interest ($27$\thinspace per cent of the analysed sample). Period
ratios with the first overtone period, $\px/\po$, fall in a range
$\sim\!(0.60, 0.64)$. Out of four analysed RRd stars none shows the
additional mode. 16 stars from the present sample were already
discovered in our previous analysis of the OGLE-III data
\citep{netzel}. Altogether, our analysis of the OGLE Galactic bulge
data led to the discovery of 262 stars of this type, to be compared
with only 23 previously known \citep[summarised in][]{pamsm15} and 18 discovered recently by \cite{jurcsik_M3}.

Basic properties of all stars are collected in Tab.~\ref{tab.list}
in the Appendix, sample of which is shown in Tab.~\ref{tab.sample}.
Subsequent columns contain period of the first overtone and of the
additional mode(s), their ratio, amplitude of the first overtone,
$\ao$, amplitude of the additional mode(s), $\ax$, and amplitude
ratio, $\ax/\ao$. Remarks are given in the last column. For some
stars there are more than one line in the table. For these stars we
detect two or three significant, well separated frequencies in the
frequency range of interest, i.e. $f/\fo\in (0.60, 0.64)^{-1}$.
Criterion for selecting the frequencies entering the table is
described in the following paragraphs.

\begin{figure}
\centering
\resizebox{\hsize}{!}{\includegraphics{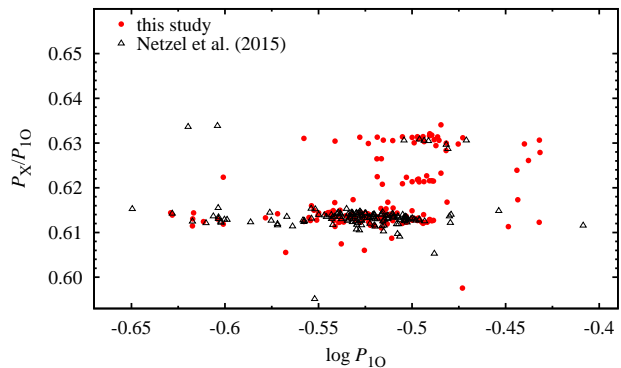}}
\caption{Petersen diagram for 0.61 stars from OGLE-IV sample analysed in this paper (red filled circles) and OGLE-III sample from our previous study \citep[triangles;][]{netzel}.}
\label{fig.pet}
\end{figure}

Data from Tab.~\ref{tab.list} are plotted in the Petersen diagram in
Fig.~\ref{fig.pet} together with the results of our study of
OGLE-III Galactic bulge data \citep{netzel}. For 20 stars more than one point
appears in the diagram (stars with more than one line in
Tab.~\ref{tab.list}).

\begin{table*} 
\centering
\caption{Sample table with properties of stars with non-radial mode (OGLE-IV). Full Table is in the Appendix.}
\label{tab.sample}
\begin{tabular}{@{}lccccccc@{}}
\hline
Name & $\po$\thinspace[d] & $\px$\thinspace[d] & $\px/\po$ & $\ao$\thinspace[mag] & $\ax$\thinspace[mag] & $\ax/\ao$ &  Remarks            \\
\hline
OGLE-BLG-RRLYR-04067 & 0.31994 & 0.19592 & 0.61236 & 0.10687 & 0.00479 & 0.0448 & a \\
OGLE-BLG-RRLYR-04105 & 0.30582 & 0.18724 & 0.61226 & 0.13009 & 0.00476 & 0.0366 &  \\
OGLE-BLG-RRLYR-04549 & 0.29964 & 0.18874 & 0.62990 & 0.12246 & 0.00384 & 0.0313 &  \\
OGLE-BLG-RRLYR-04599 & 0.28939 & 0.17797 & 0.61498 & 0.13980 & 0.00340 & 0.0243 & a,g \\
OGLE-BLG-RRLYR-04754 & 0.28631 & 0.17578 & 0.61394 & 0.12788 & 0.00453 & 0.0354 & g,h \\
OGLE-BLG-RRLYR-04762 & 0.29465 & 0.18061 & 0.61295 & 0.12640 & 0.00700 & 0.0554 &  \\
\multicolumn{8}{c}{$\dots$}\\
\hline
\end{tabular}
\end{table*}

\begin{figure}
\centering
\resizebox{\hsize}{!}{\includegraphics{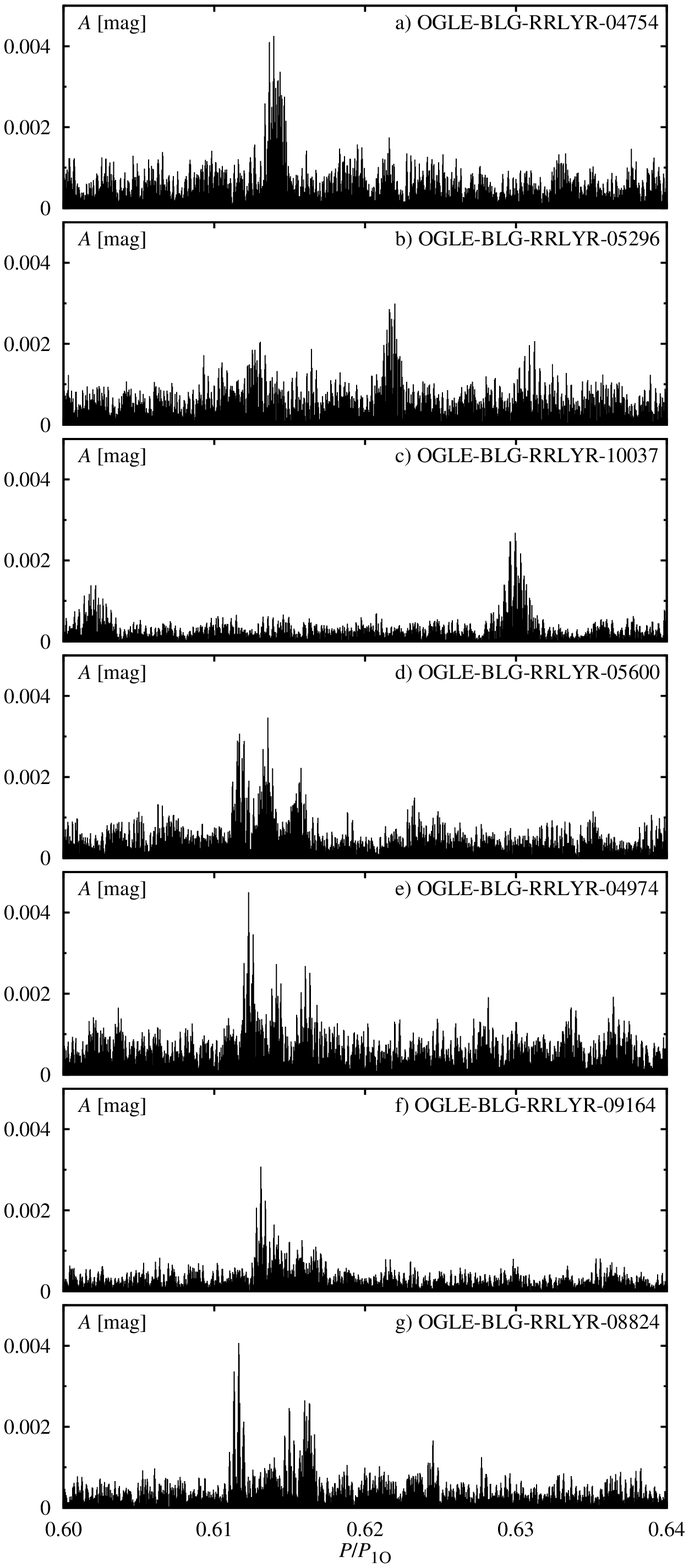}}
\caption{Examples of signal structures detected in the frequency
         spectra of 6 stars at frequency range characteristic for
         additional modes discussed in this paper. Note, that
         horizontal axis is a normalized period scale, ie.,
         $P/\po=\fo/f$.}
\label{fig.struct}
\end{figure}

In case of many stars there are many resolved peaks with $S/N\geq 4$
in the frequency range of interest. Examples are presented in
Fig.~\ref{fig.struct}. Note that in this and other figures
presenting frequency spectra, we use a more convenient, normalized
period scale for horizontal axis, i.e. we plot frequency spectra vs.
$P/\po=\fo/f$. This allows a direct comparison with the Petersen
diagram. Usually the signals appear as a relatively narrow cluster
of peaks centered at a characteristic frequency. Panel a of
Fig.~\ref{fig.struct} shows OGLE-BLG-RRLYR-04754, which displays
such cluster of peaks centered at $P/\po\approx0.613$. Next two
panels show analogous clusters of peaks, but centered around $0.622$
(panel b) and $0.631$ (panel c). In case of these and similar stars
only the frequency and amplitude of the highest peak in a cluster is
included in Tab.~\ref{tab.list}, regardless whether other signals
from the cluster have $S/N\geq 4$ (before or after prewhitening).
Those stars are marked with `g' in the remarks column of
Tab.~\ref{tab.list}.

Such structures, with signals in three different, rather
characteristic locations, are most commonly found in the frequency
spectra of the analysed stars. As a result three sequences in the
Petersen diagram emerge (Fig.~\ref{fig.pet}). Most of the stars are
placed in the lowest sequence. It is horizontal and centered at
$\px/\po\approx 0.613$ value. It also covers the largest range of
periods. Less populated sequence is the highest one. It is centered
at $\px/\po\approx 0.631$. A slight trend of decreasing period ratio
with increasing period of the first overtone is visible in this
sequence. A third sequence is weakly populated but well visible in
Fig.~\ref{fig.pet}, in between the two previously described
sequences. It appears horizontal, is centered at $\px/\po\approx 0.623$ and covers mostly
a small range of periods around $\log\po=-0.5$. Although stars which fit this sequence were also detected in previous studies (4 stars in \cite{jurcsik_M3} and 4 stars from tab.~8 in \cite{pamsm15}), the fact, that they form a third sequence became evident only in the OGLE-IV data. The three sequences
are almost equidistant in the Petersen diagram. A significant spread of
period ratios is present within each sequence. It is a consequence
of the just described form in which signals appear in the frequency
spectrum. The signals in a cluster of peaks are of almost equal
height (Fig.~\ref{fig.struct}). Which peak is the highest and hence
included in Tab.~\ref{tab.list}/Fig.~\ref{fig.pet} is a matter of
chance, taking into account the relatively small $S/N$ for these
peaks, typically around $5-6$.

Although this choice is, out of necessity, a bit arbitrary, based on
Fig.~\ref{fig.pet} and using data only for the stars detected in
homogeneous OGLE Galactic bulge samples, we can estimate the values
of period ratio separating the three sequences. These fall around
$P_{\rm x}/P_{\rm 1O}\approx 0.620$ and $P_{\rm x}/P_{\rm
1O}\approx 0.628$. The three sequences selected this way are named
0.61-, 0.62- and 0.63-sequences in the following.

In several stars, the clusters of peaks seem to form triplets. These
stars are marked with `t' in the remarks column of
Tab.~\ref{tab.list}. Panel d in Fig.~\ref{fig.struct} shows an
example of such triplet visible in a spectrum of
OGLE-BLG-RRLYR-05600, which belongs to 0.61 sequence. The highest
peak is detected in the central component of the triplet and its
frequency/amplitude are included in Tab.~\ref{tab.list}.
 It may also happen that the highest peak in the triplet
is detected in its side component. This is the case for
OGLE-BLG-RRLYR-04974 (panel e in Fig.~\ref{fig.struct}), another
star from 0.61 sequence. Its frequency is then reported in
Tab.~\ref{tab.list}/Fig.~\ref{fig.pet}. The spread of period ratios
within 0.61 sequence (Fig.~\ref{fig.pet}) is partially due to the
appearance of such triplets with various heights of the triplet
components. We also note that the separation between the triplet components 
(their midpoints) is approximately equal  (see also Sect.~\ref{ssec:comp} 
and \ref{sec:summary}). 
In a few stars we observe doublets instead of triplets.
These stars are marked with `d' in the remarks column of
Tab.~\ref{tab.list}. We note that triplets and doublets appear 
preferentially in the lowest, 0.61 sequence, and only rarely in 
the two others.

Besides single clusters of peaks (panels a, b and c in
Fig.~\ref{fig.struct}), and multiplet-like structures (panels d and
e), there are more complex structures which are not easily
classifiable (`f' in remarks column). Two examples are presented in
panels f and g in Fig.~\ref{fig.struct}: OGLE-BLG-RRLYR-09164 and
OGLE-BLG-RRLYR-08824. In case of the first star, there is a power
excess covering a rather wide frequency range. The second star has
several signals between 0.61 and 0.62, which do not form a
triplet-like structure. In all those stars the highest peak is
chosen for the table and the Petersen diagram. 
The complex structures appear only in the lowest, 0.61 sequence.

\begin{figure}
\centering
\resizebox{\hsize}{!}{\includegraphics{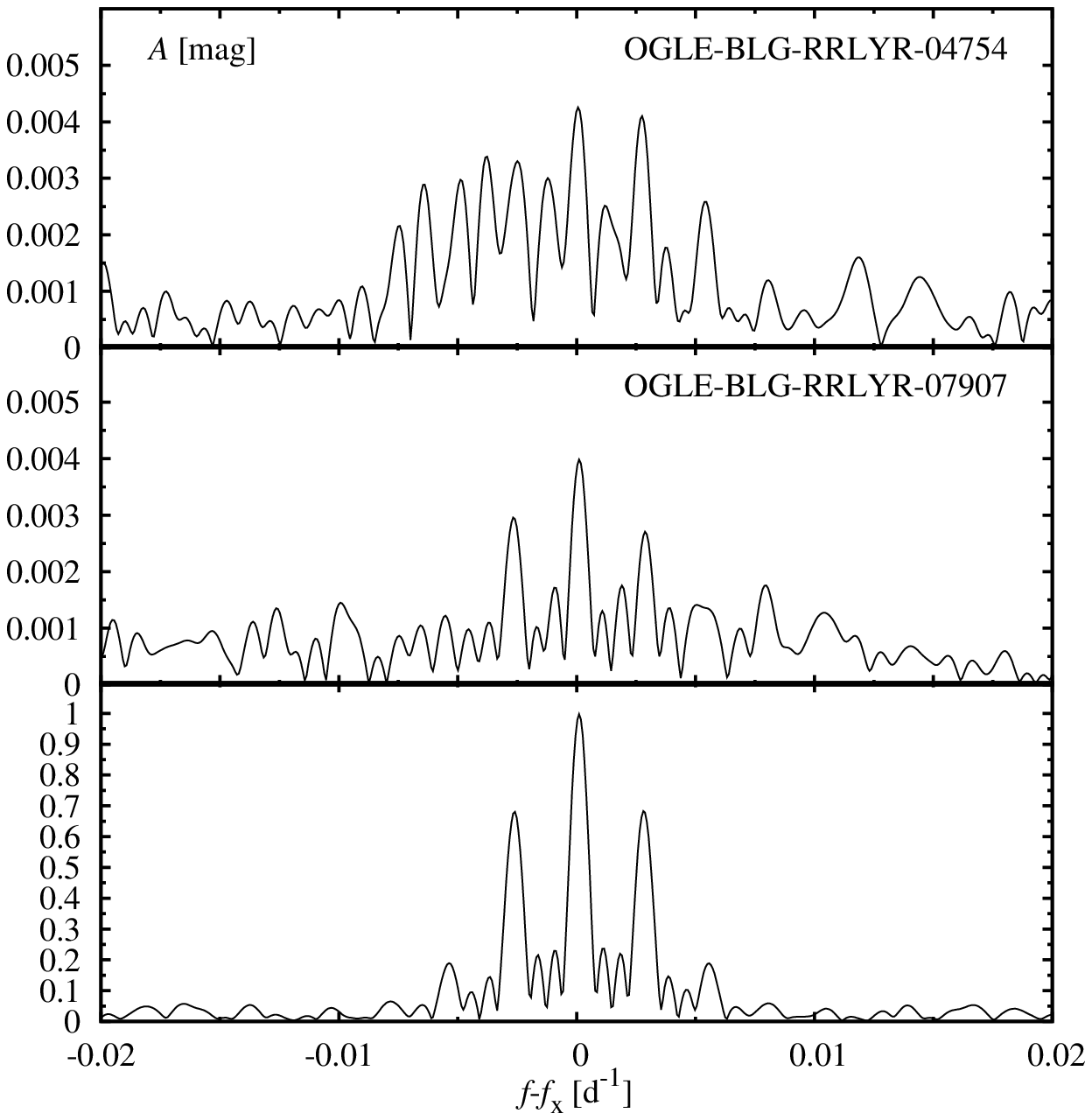}}
\caption{Comparison of signal at $\fx$ for two stars (two upper panels) with spectral window (bottom panel).}
\label{fig.win-por}
\end{figure}

To get more insight into appearance of the discussed complex structures in the frequency spectra, in Fig.~\ref{fig.win-por} we present a comparison between spectral window and $\fx$ signal in two stars. Middle panel of this figure shows spectrum of OGLE-BLG-RRLYR-07907 centered at $\fx$. In this star additional signal is discrete and can be substracted from the data with a single sine function. Structure is similar to the spectral window shown on the bottom panel. Upper panel in Fig.~\ref{fig.win-por} shows the additional mode in OGLE-BLG-RRLYR-04754 (this star is also shown in panel `a' of Fig.~\ref{fig.struct} in wider frequency range). The mode appears as a cluster of peaks rather than a single and coherent peak.

Similar structures to described above were also detected in the {\it
Kepler} RRc data \citep[e.g. Fig.~12 in][see also
Section~\ref{ssec:comp}]{pamsm15}.

\begin{figure}
\centering
\resizebox{\hsize}{!}{\includegraphics{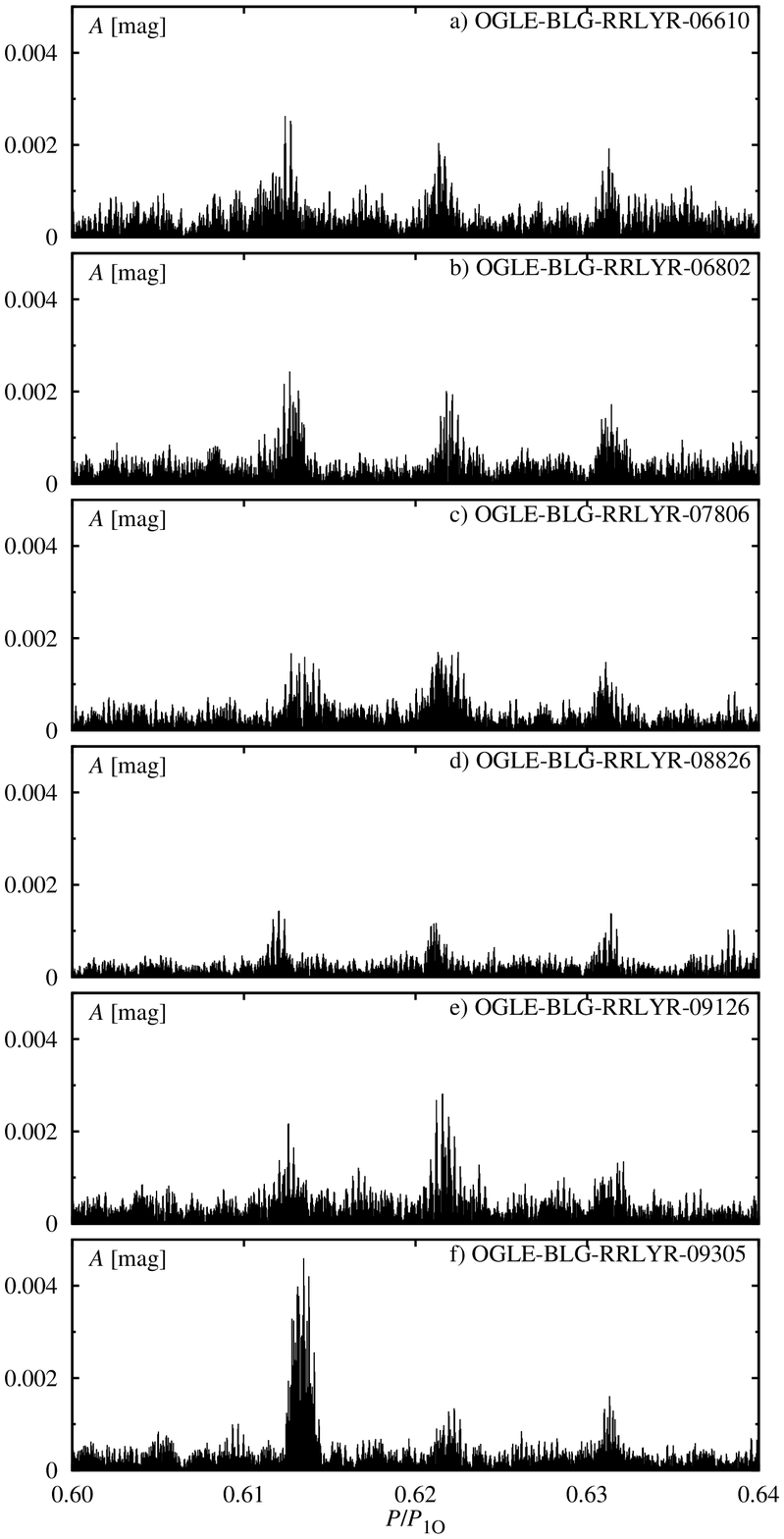}}
\caption{Frequency spectra for 6 stars in which we detect
         significant peaks corresponding to three different
         sequences in the Petersen diagram, Fig.~\ref{fig.pet}.}
\label{fig.ts}
\end{figure}

\begin{figure}
\centering
\resizebox{\hsize}{!}{\includegraphics{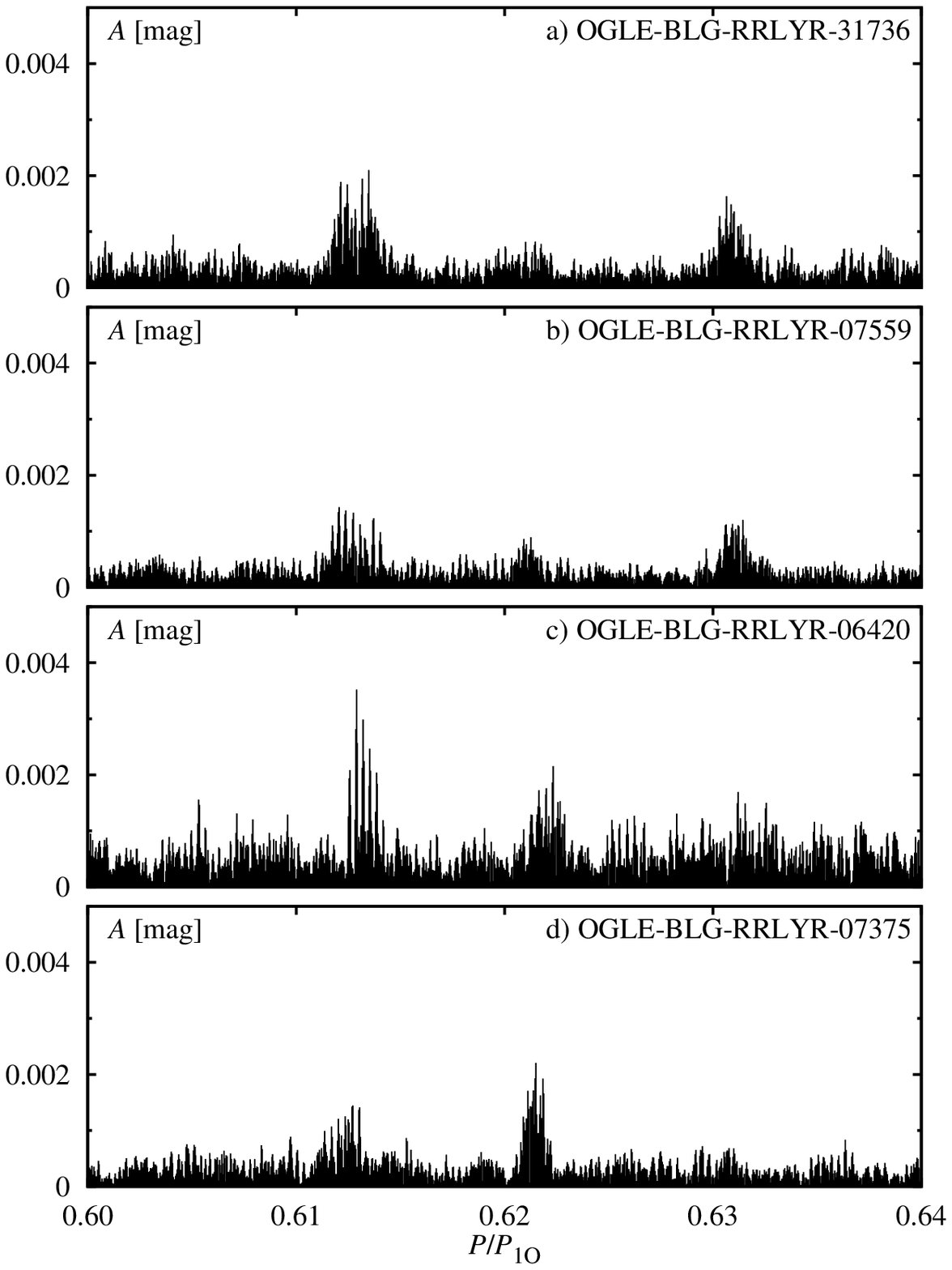}}
\caption{Frequency spectra for stars in which we detect significant
         signals corresponding to two of the three sequences in the
         Petersen diagram, Fig.~\ref{fig.pet}.}
\label{fig.2s}
\end{figure}

Some stars are important exceptions from the scenario outlined above
and detected in space observations of RRc stars. In
Fig.~\ref{fig.ts} we present frequency spectra for six stars in
which we observe simultaneously three signals falling into three
sequences defined above. Again, typically, we do not observe three
single peaks, but three clusters of peaks centered at frequencies
characteristic for the three sequences. For these stars, data for
the highest peak in each of the three clusters are given in
Tab.~\ref{tab.list} (three rows) and in the Petersen diagram
(Fig.~\ref{fig.pet}) each of these stars is represented by three
points. We also find stars with two clusters of peaks corresponding
to two of the three sequences we defined. Examples are shown in
Fig.~\ref{fig.2s}. For these stars, data for two highest peaks, one
from each cluster, are given in Tab.~\ref{tab.list} and two points
are plotted in the Petersen diagram.

We also note that in several stars in which only one additional peak
belonging to one of the sequences is present, we observe a power
excess centered at frequency characteristic for other sequences.
Since these bumps of power excess are below $S/N$ of 4 no additional
information is included in Tab.~\ref{tab.list} and such stars are
represented in the Petersen diagram with only one point.

\begin{figure}
\centering
\resizebox{\hsize}{!}{\includegraphics{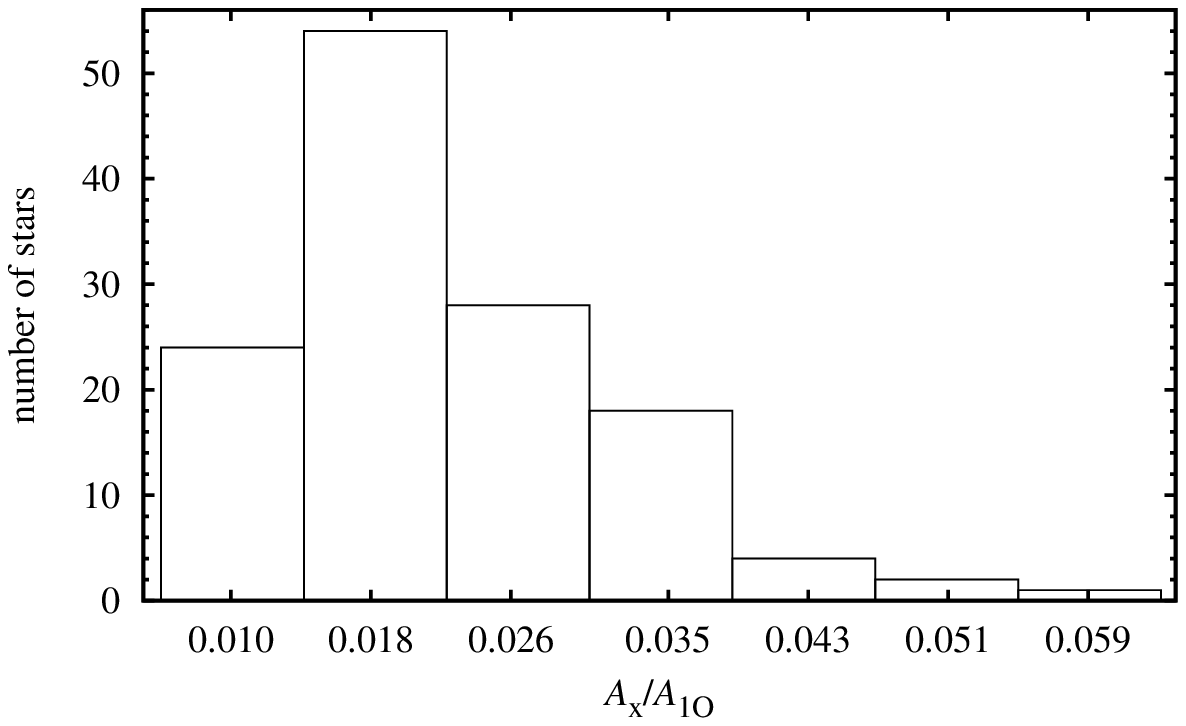}}
\caption{Histogram of amplitude ratios, $\ax/\ao$. One signal for each star included.}
\label{fig.ar}
\end{figure}

Amplitudes of the additional modes are very low compared to
amplitudes of the first overtone mode.  Histogram of amplitude
ratio, $\ax/\ao$, for all analysed stars is shown in
Fig.~\ref{fig.ar}. Amplitudes of the additional modes vary from
$0.6$ per cent to $5.5$ per cent of the first overtone amplitude
with average value of $2$ per cent.

\begin{figure}
\centering
\resizebox{\hsize}{!}{\includegraphics{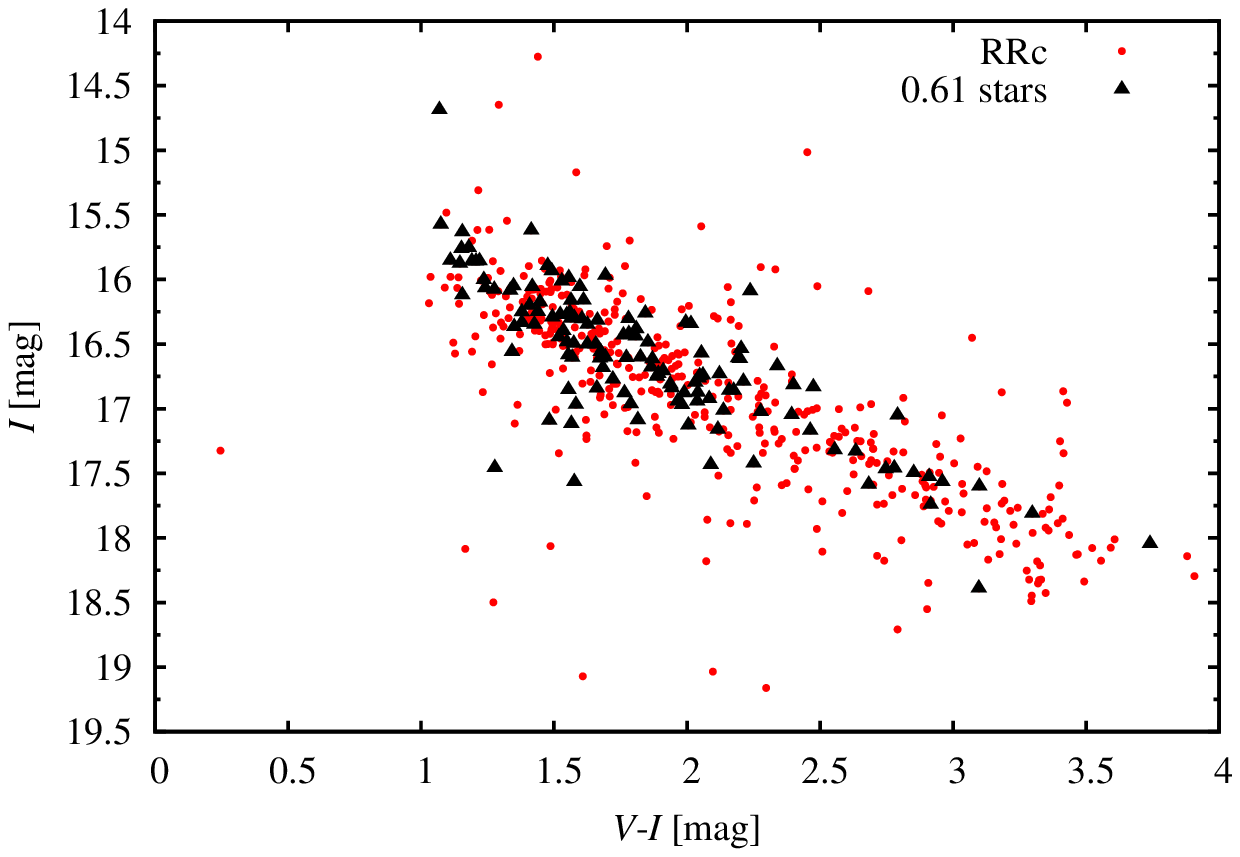}}
\caption{The observed color-magnitude diagram for RRc stars in the two analysed OGLE-IV fields.}
\label{fig.cmd}
\end{figure}

In Fig.~\ref{fig.cmd} we present the color-magnitude diagram for RRc stars in the two analysed OGLE-IV fields. Stars with the additional mode follow a progression of single-periodic RRc stars and are not restricted to any particular color range in the diagram. We note however, that the large extent of the progression is caused by the different reddenings towards the observed stars, which is characteristic for the extended Galactic bulge population.

\begin{figure}
\centering
\resizebox{\hsize}{!}{\includegraphics{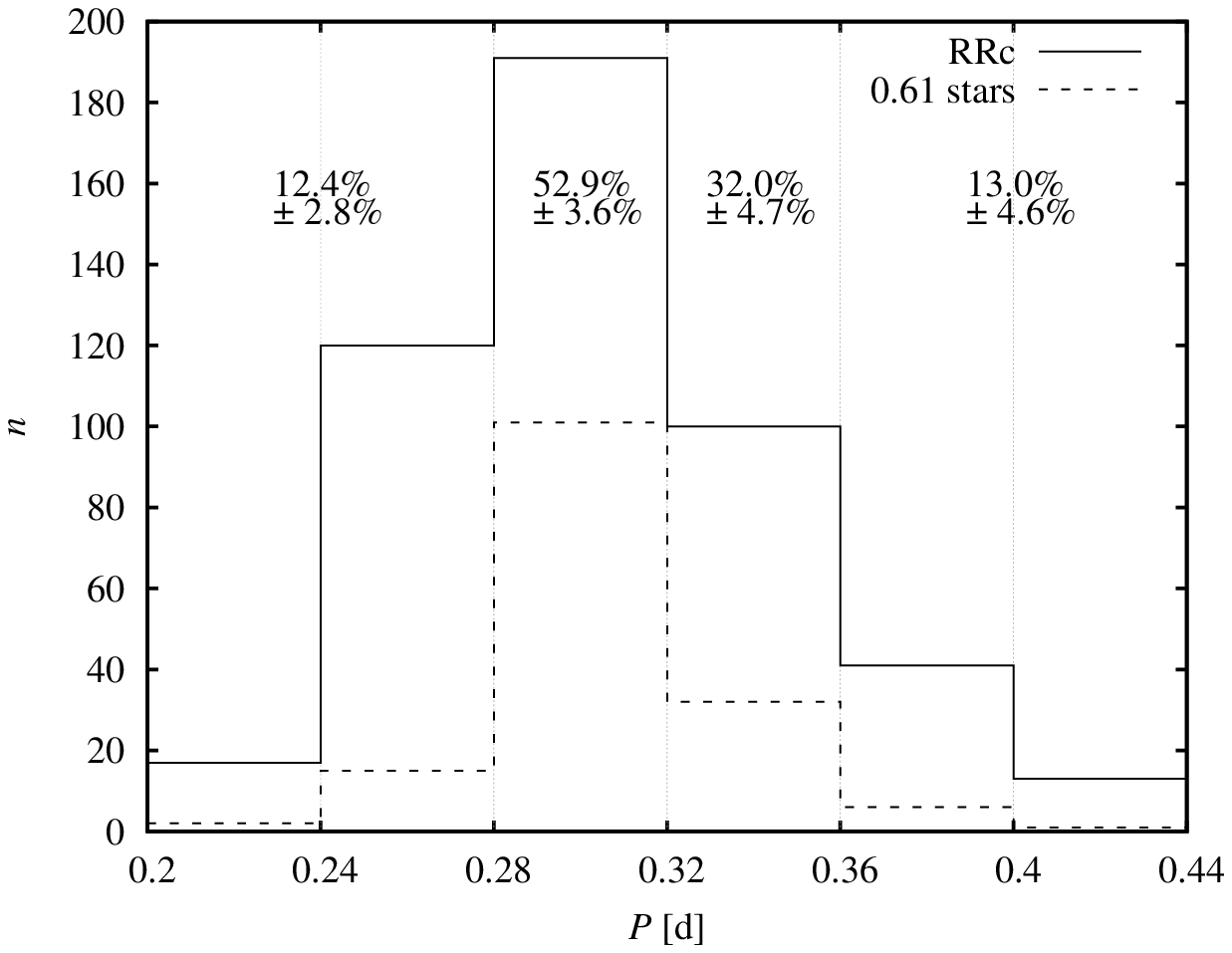}}
\caption{Histogram of periods, $P_{1{\rm O}}$, for RRc sample and 0.61 stars. Incidence rate for 0.61 stars is calculated in four bins. Outer bins are merged, because of a small number of stars. Statistical errors of the incidence rates were calculated assuming that the population follows a Poisson distribution \citep[e.g.][]{alcock}.}
\label{fig.hist_p}
\end{figure}

In Fig.~\ref{fig.hist_p} we show the histogram of first overtone period distribution for RRc stars in the two analysed fields. The 0.61 stars are not confined to any particular period range. The distribution peaks in between 0.28 and 0.32 d, but both long and short period 0.61 stars are also present. Together with the color-magnitude plot, it suggests that 0.61 stars are not confined to any particular part of the HR diagram, but are distributed over entire first overtone instability strip \citep[cf. with][]{jurcsik_M3}.

16 stars, which were discovered earlier in OGLE-III data \citep{netzel}, are marked in Tab.~\ref{tab.list} with `h' in remarks. There is no significant difference between properties of these stars determined from OGLE-III and from OGLE-IV data. For a few stars, in OGLE-IV data we detected subharmonics and combination frequencies of the first-overtone and of the additional mode, which were undetected in OGLE-III data. In three stars we detected frequencies belonging to more than one of the three sequences described earlier. Due to high observing cadence, the noise level in OGLE-IV data is much lower, which makes detection of other, low-amplitude signals possible.

\subsection{Seasonal variations}\label{ssec:seasonal}

\begin{figure}
\centering
\resizebox{\hsize}{!}{\includegraphics{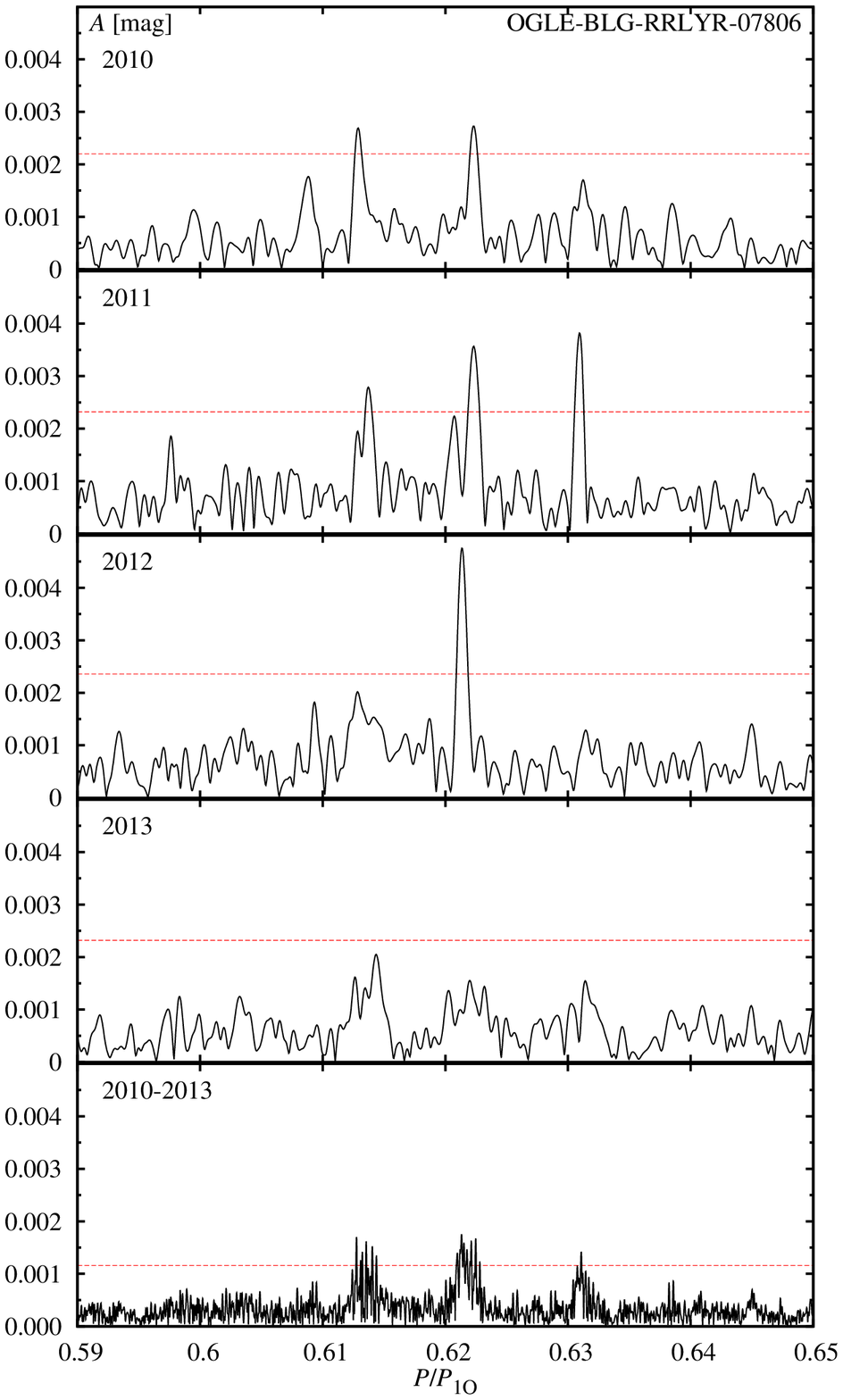}}
\caption{Frequency spectra for each of the four observation seasons
         (first four panels) for OGLE-BLG-RRLYR-07806 at the
         frequency range of interest. Last panel shows frequency
         spectrum of all the data. Horizontal line indicates the $4\sigma$ noise level.}
\label{fig.07806-sezony}
\end{figure}

Complex structures observed in the frequency spectrum at the
frequencies of the additional modes and at their vicinity indicate strong
non-stationarity of the modes. In order to investigate the variation
of the additional signals, we analysed separately each of the four
observing seasons. Example of seasonal analysis is shown in
Fig.~\ref{fig.07806-sezony} for OGLE-BLG-RRLYR-07806. First four
panels correspond to subsequent observing seasons. Last panel shows
the frequency spectrum of all data analysed together. The green
dotted line indicates $S/N=4$. Analysis of the whole dataset shows
three clusters of signals corresponding to the three sequences on
the Petersen diagram, with $P_{\rm x}/P_{\rm 1O}\sim 0.61$, $\sim
0.62$ and $\sim 0.63$.

Analysis of each season separately shows strong variability within
each cluster. In the first observing season, peaks from the two
sequences, 0.61 and 0.62, are significant. In the second observing
season we clearly detect three peaks, each corresponding to one of
the sequences. Signal corresponding to 0.63 sequence is the highest.
In the third observing season there is only one signal from 0.62
sequence. Analysis of the fourth observing season does not reveal
any significant frequencies in the range of interest, however power
excess is obvious at the expected locations. Amplitudes and
frequencies of the observed signals clearly vary, producing, in the
frequency spectrum of all data, three clusters of peaks, rather than
three coherent peaks.

Seasonal changes described above are characteristic for most stars. Frequency and amplitude of the additional mode vary from season to season in every star. Changes are irregular \citep[for an example see also fig. 9 in][]{netzel}. Also, signals corresponding to different sequences vary differently, which is well visible in Fig.~\ref{fig.07806-sezony}.

We note that the OGLE data do not allow more detailed analysis than on a season-to-season basis. On the other hand, we know, at least from {\it Kepler} observations of this type of stars \citep{pamsm15}, and the appearance of the frequency spectra, that the variation may occur on a much shorter time scale, which is averaged out in our analysis.

Seasonal analysis discussed above indicates that the additional non-radial modes could be 
excited in other RRc stars of our sample, not listed in
Tab.~\ref{tab.list}. One can imagine that the signals are strong
only during one or two seasons, but in the analysis of the whole
dataset they just manifest as power excess at expected frequencies,
but below the adopted detection threshold ($S/N=4$). This is indeed the case. In some stars, presented in Fig.~\ref{fig.2s}, with significant signals belonging to two of the three sequences, a power excess at the location expected for the remaining, third sequence, is also present, but is too low to be considered significant. Consequently it was not included in the Tab.~\ref{tab.list}.

\clearpage

\subsection{Subharmonic frequency range}\label{ssec:subh}

Significant signals at subharmonics of the additional mode are
commonly detected in RRc stars observed from space, both at $1/2\fx$
and at $3/2\fx$ \citep{szabo_corot,pamsm15,molnar}. In the ground based observations
subharmonics are hard to detect. Only in \cite{netzel} we reported a
weak signals at $1/2\fx$ in four stars.

In OGLE-IV data we detect significant signals at around subharmonic
of additional mode(s), i.e. at around $1/2\fx$, in 26 stars. These
stars are marked with `s' in the remarks column of
Tab.~\ref{tab.list}. Only in two stars we detect signal at around
$3/2\fx$ with $S/N>4$. Strictly, we do not detect single peaks
exactly at $1/2\fx$ or at $3/2\fx$. Typically we detect a wide power
excesses at around $1/2\fx$. The structures observed in subharmonic
frequency range are as complex, and even more diverse, than the just
described structures at $f_x$ and its vicinity (parent frequency
range, in the following).

\begin{figure*}
\centering
\resizebox{0.32\hsize}{!}{\includegraphics{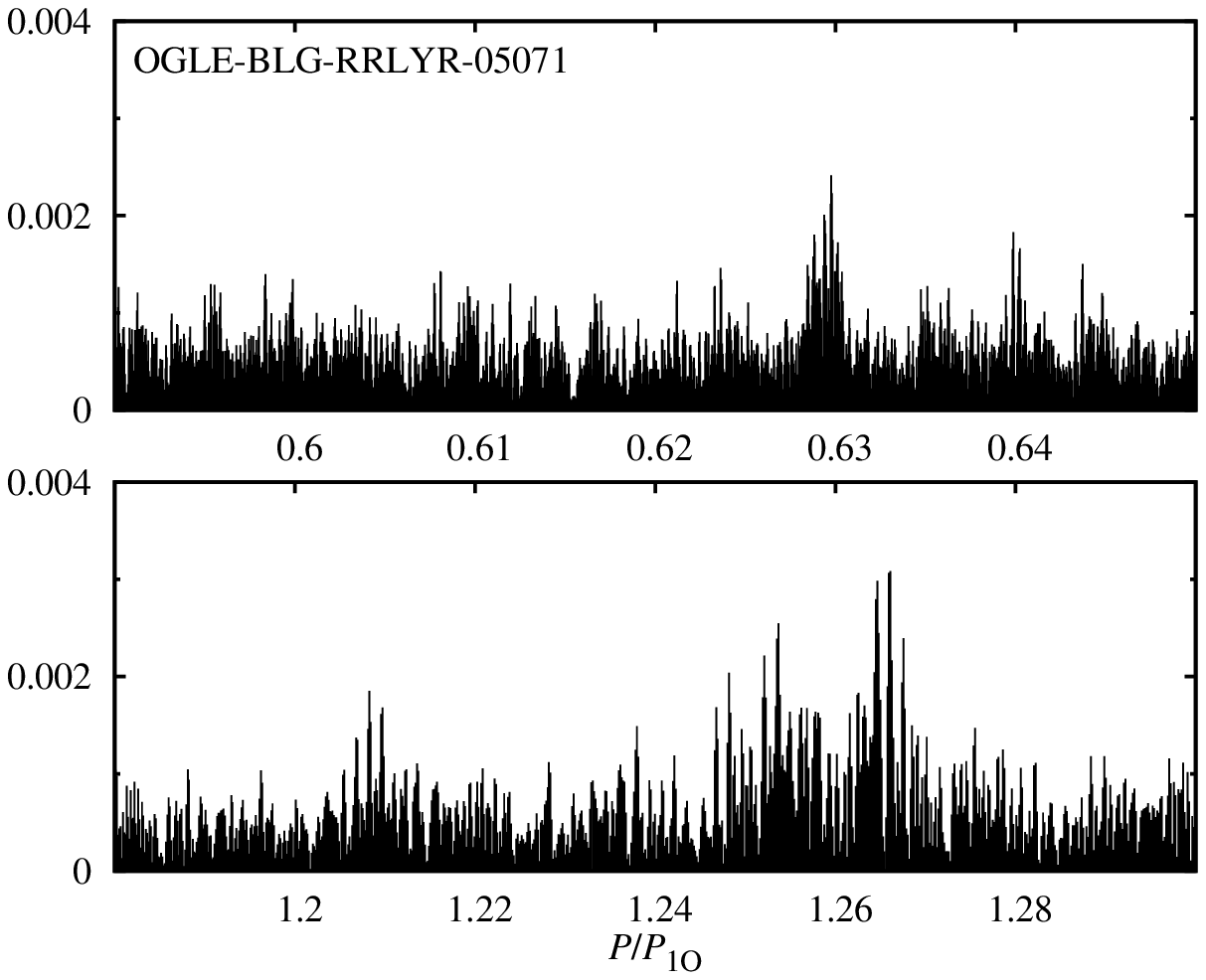}}
\resizebox{0.32\hsize}{!}{\includegraphics{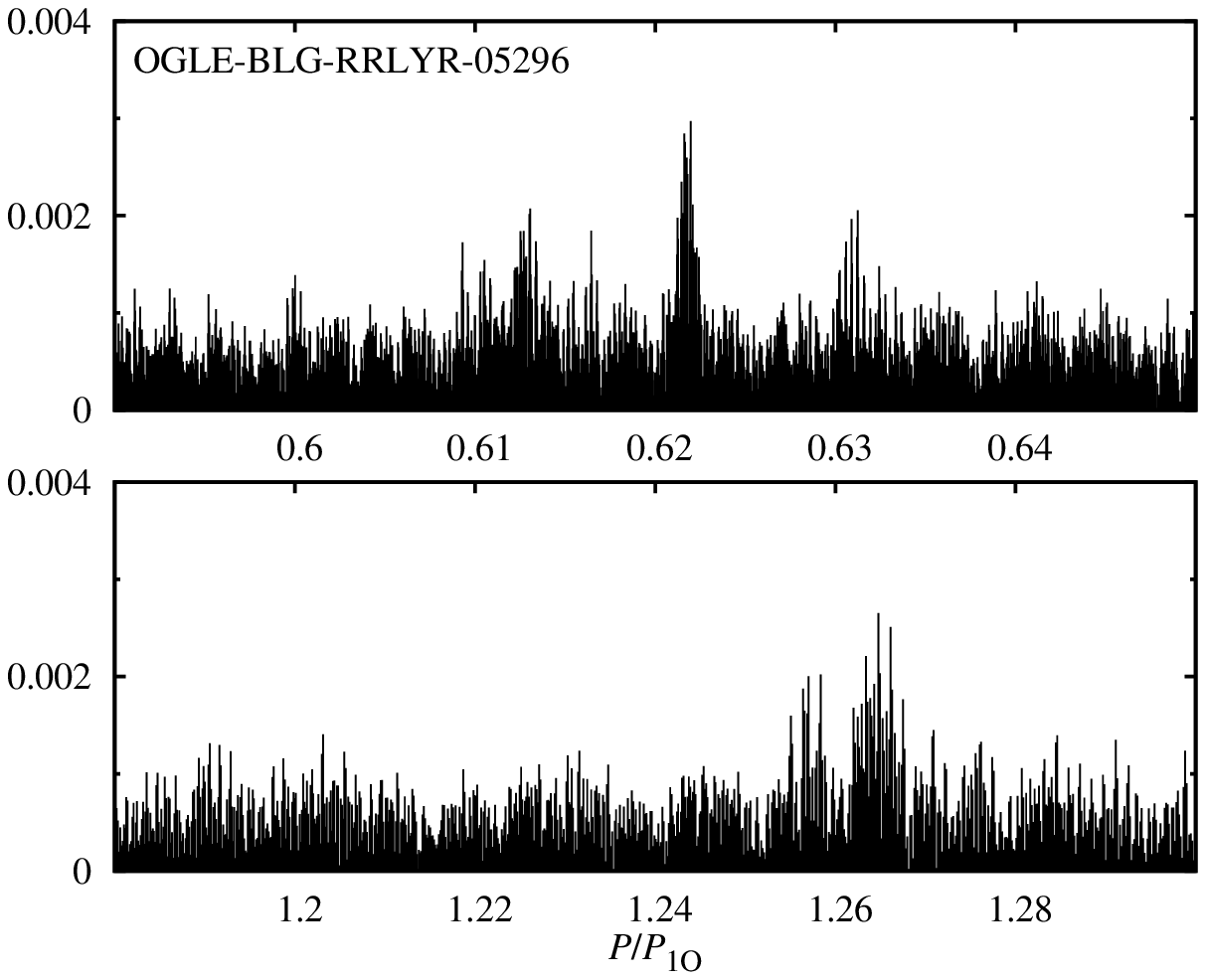}}
\resizebox{0.32\hsize}{!}{\includegraphics{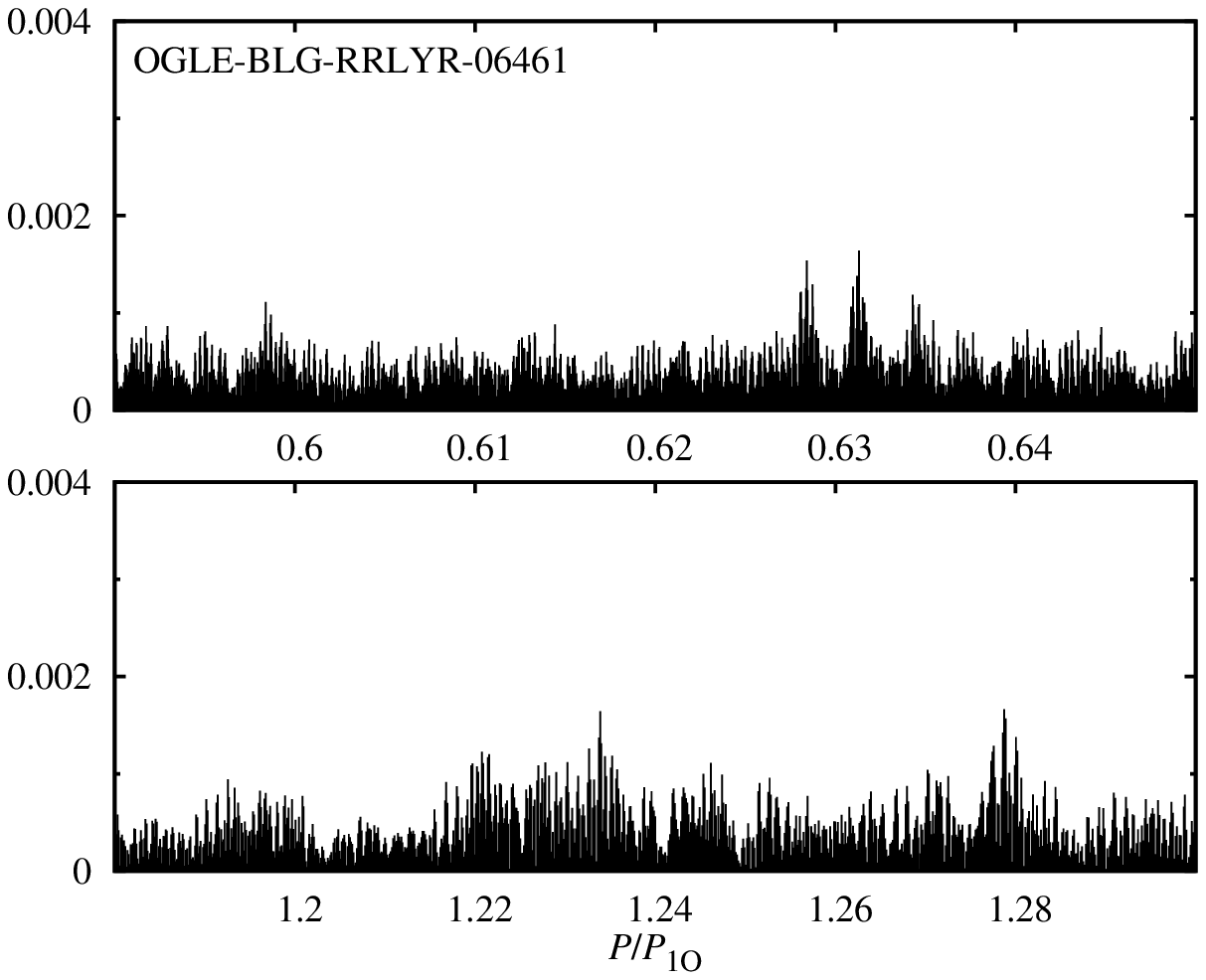}}\\
\vspace*{.2cm}
\resizebox{0.32\hsize}{!}{\includegraphics{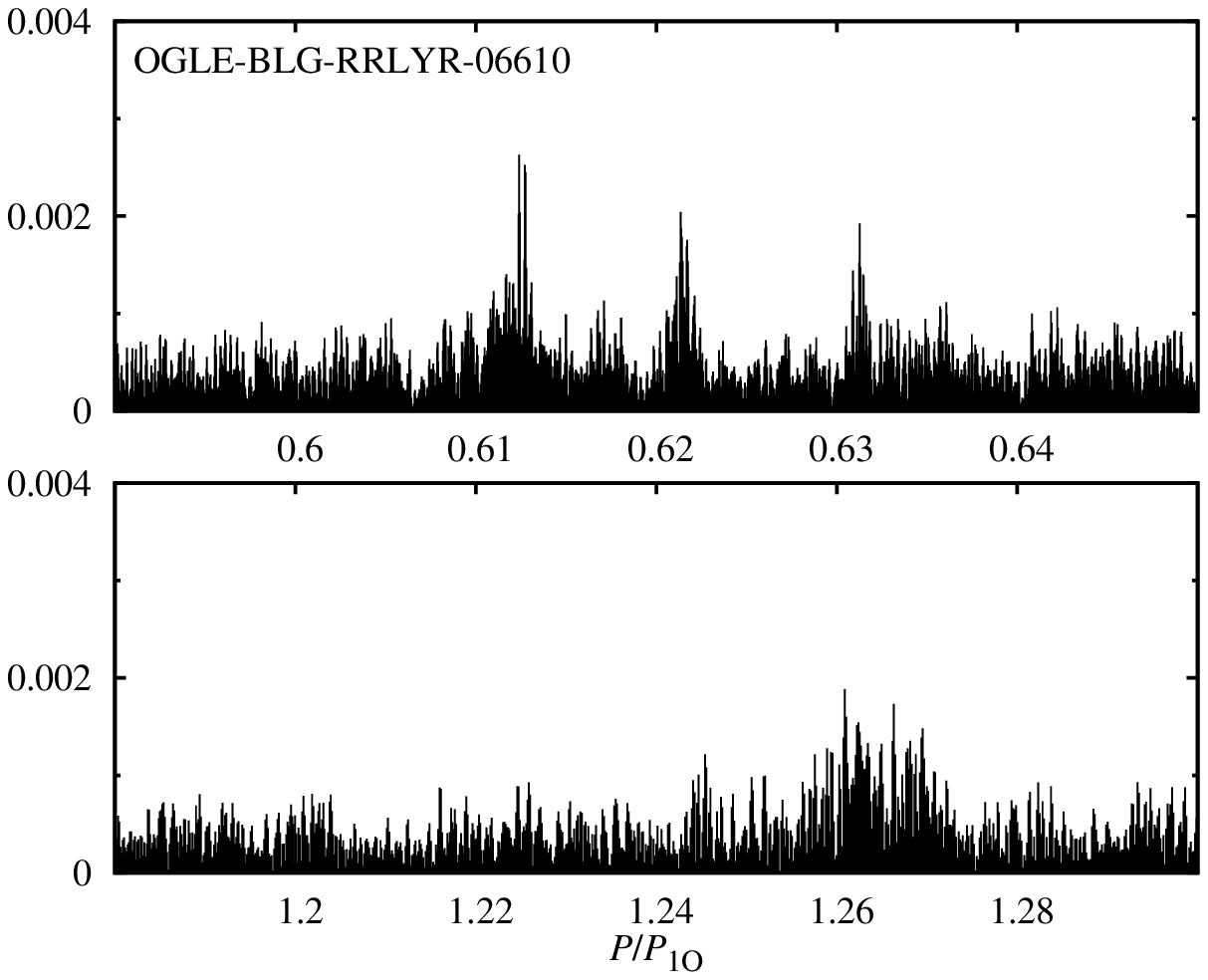}}
\resizebox{0.32\hsize}{!}{\includegraphics{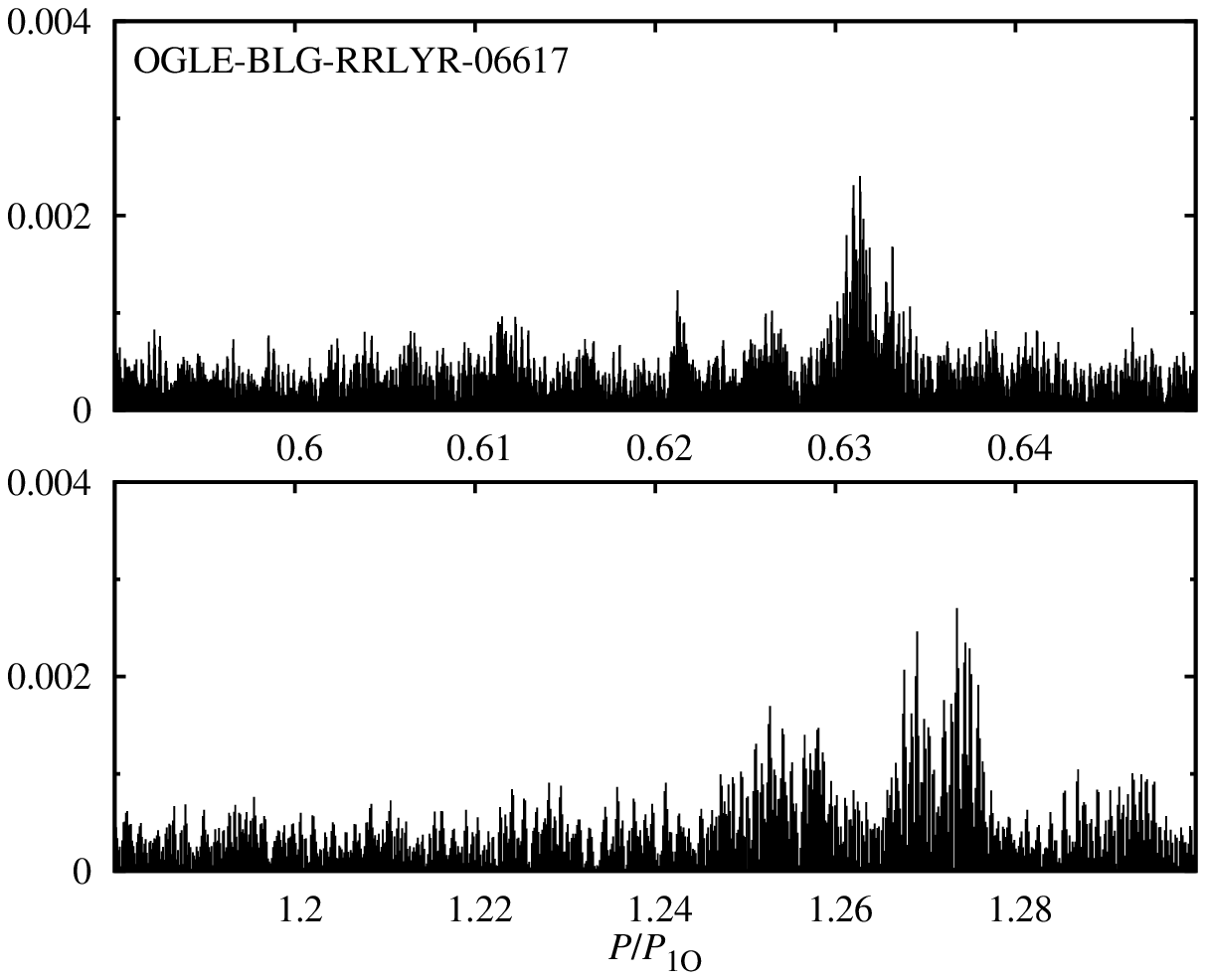}}
\resizebox{0.32\hsize}{!}{\includegraphics{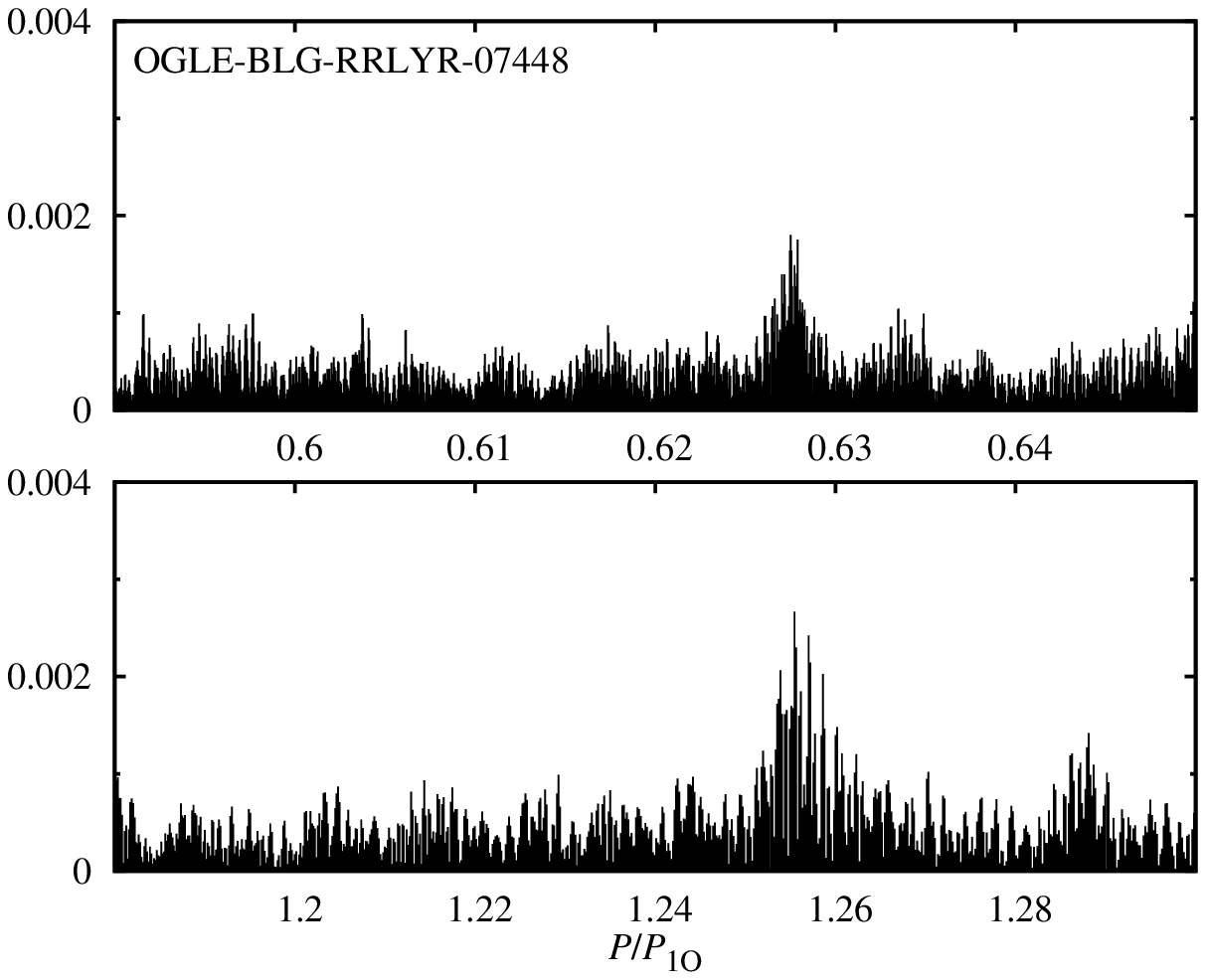}}\\
\vspace*{.2cm}
\resizebox{0.32\hsize}{!}{\includegraphics{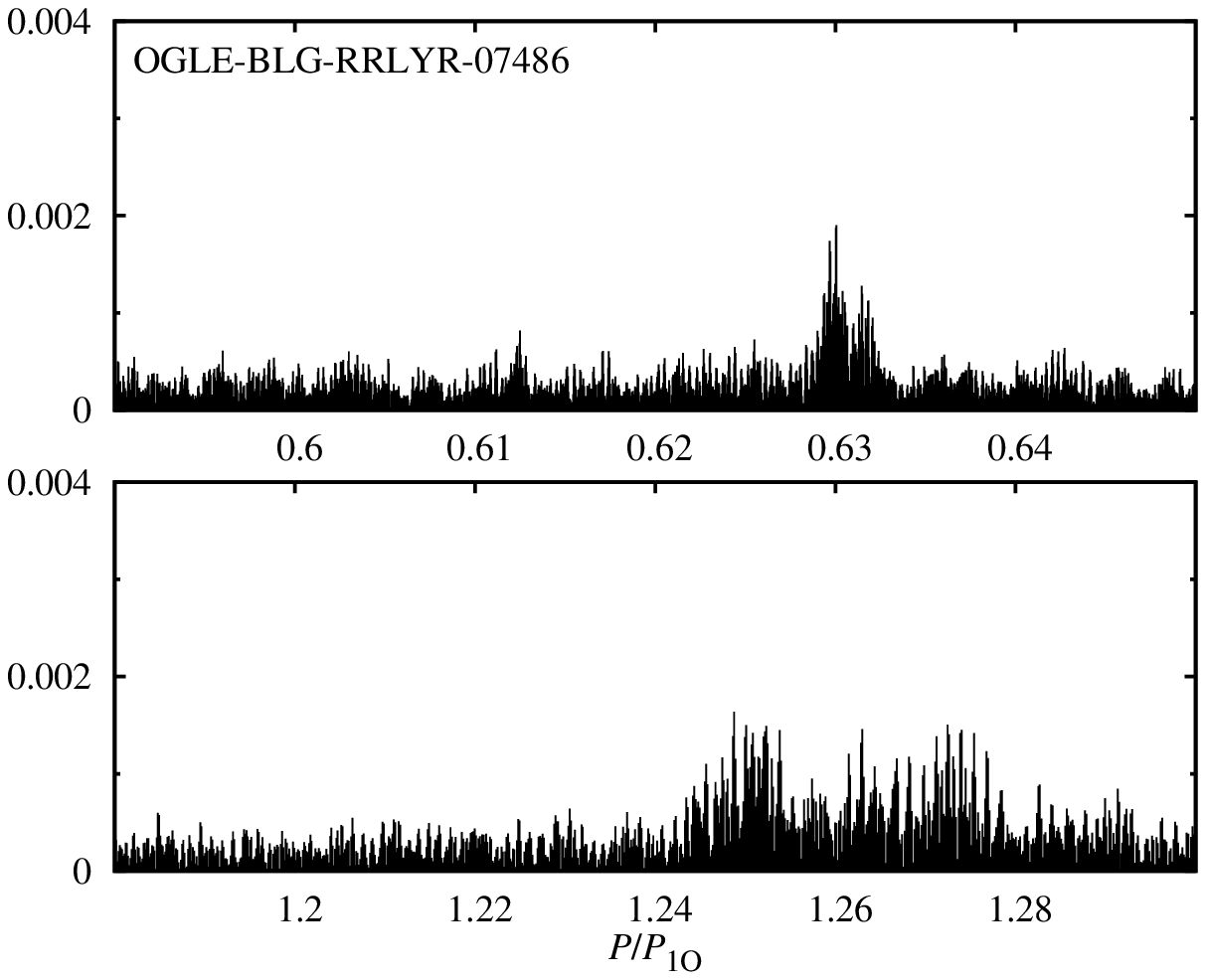}}
\resizebox{0.32\hsize}{!}{\includegraphics{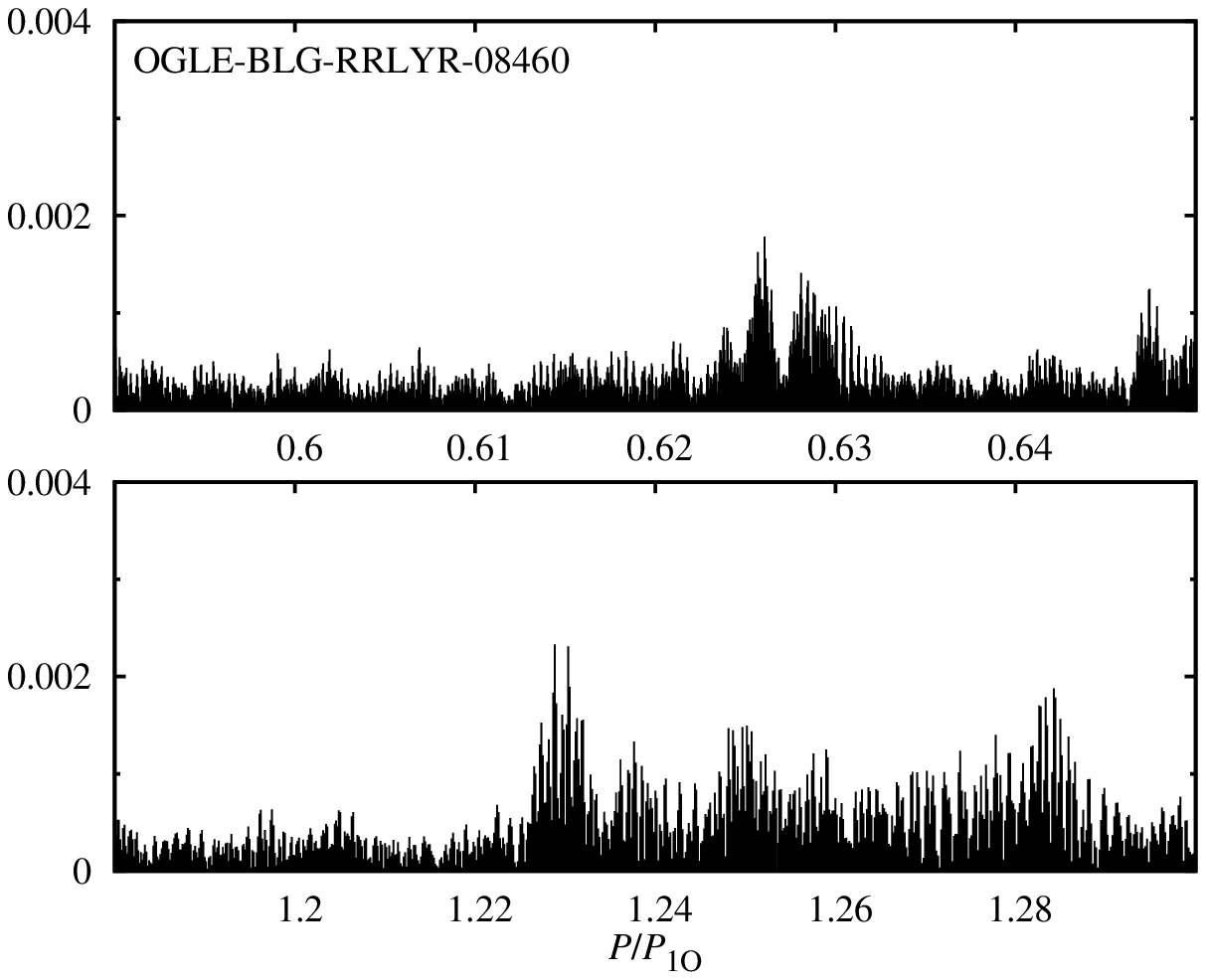}}
\resizebox{0.32\hsize}{!}{\includegraphics{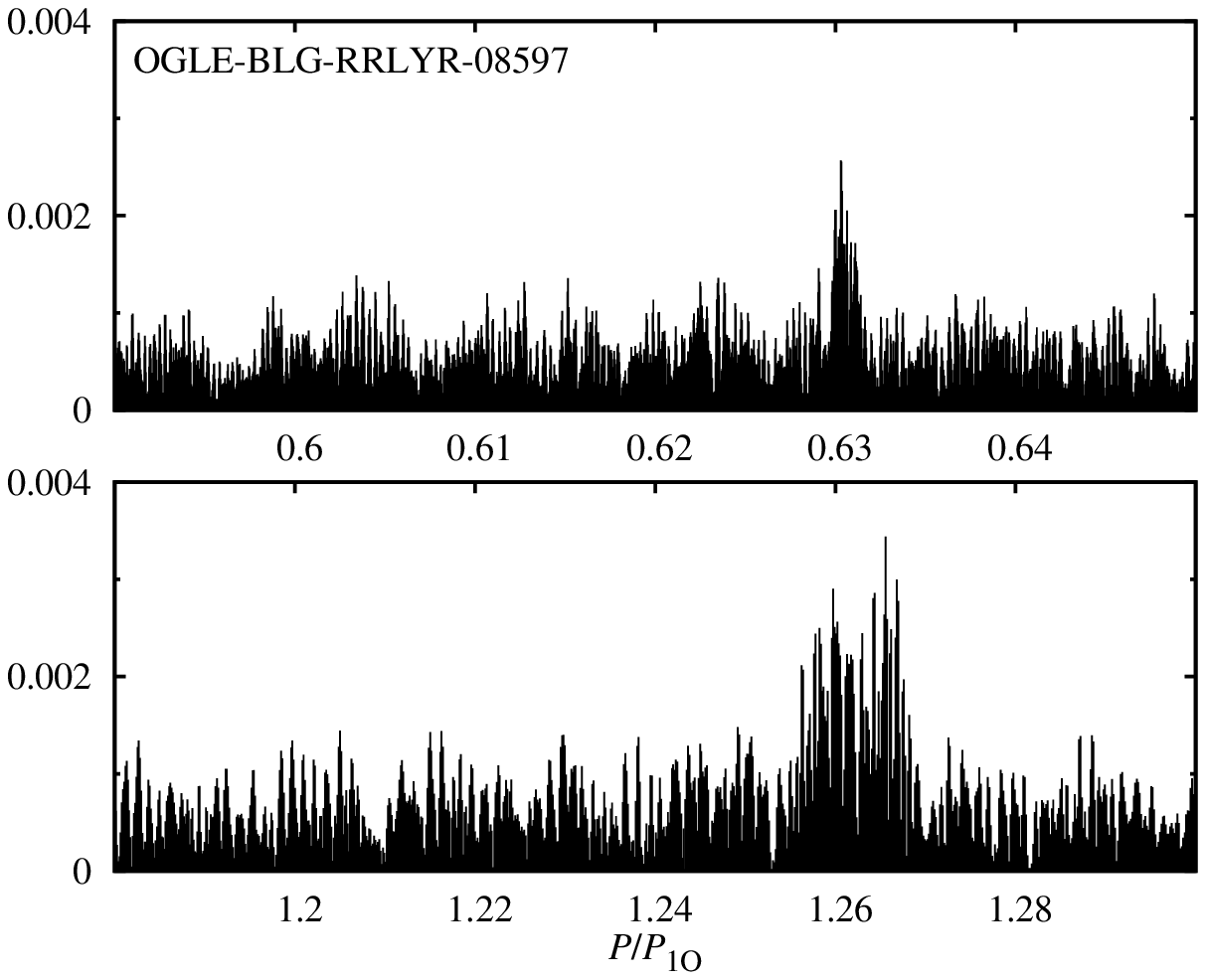}}\\
\vspace*{.2cm}
\resizebox{0.32\hsize}{!}{\includegraphics{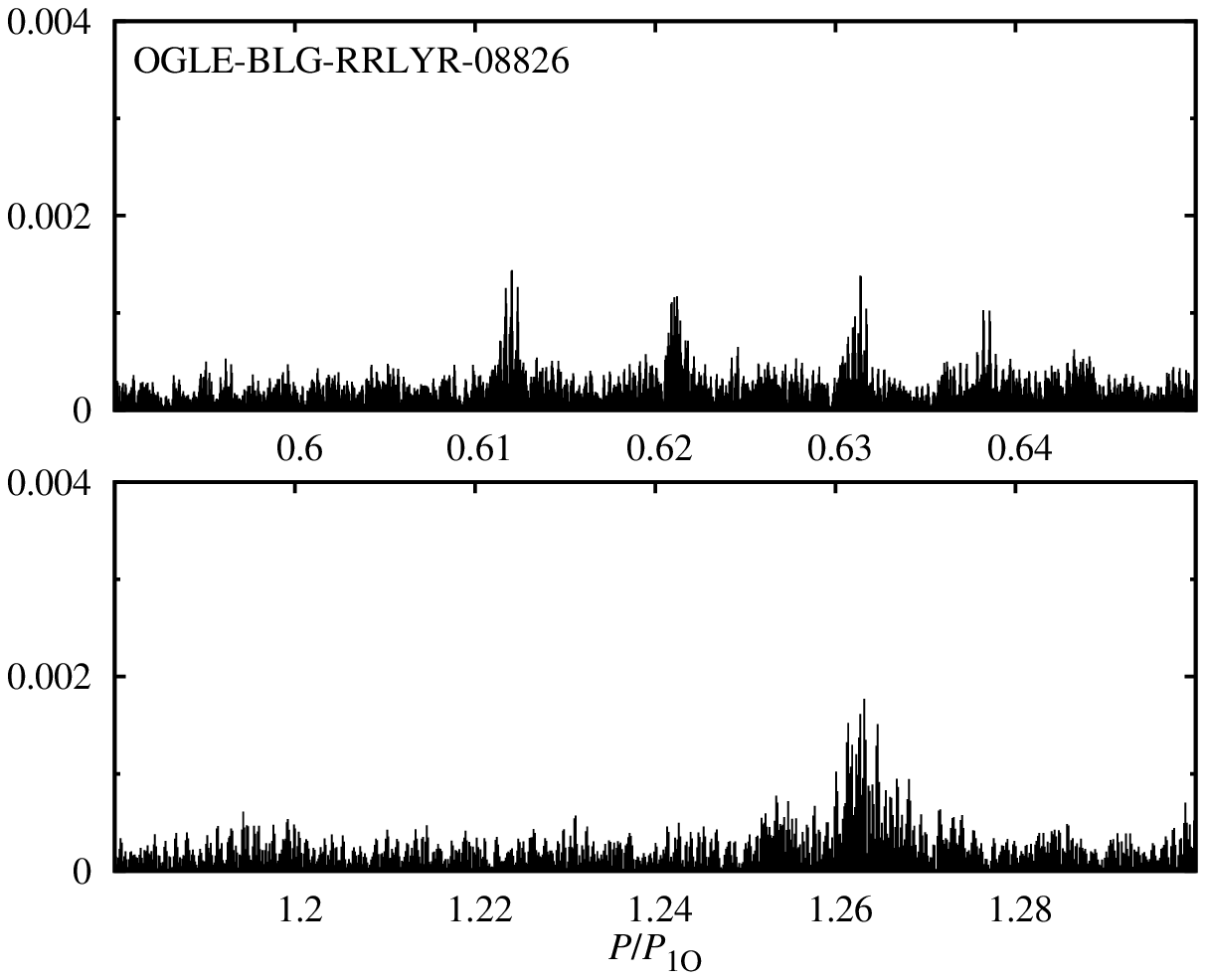}}
\resizebox{0.32\hsize}{!}{\includegraphics{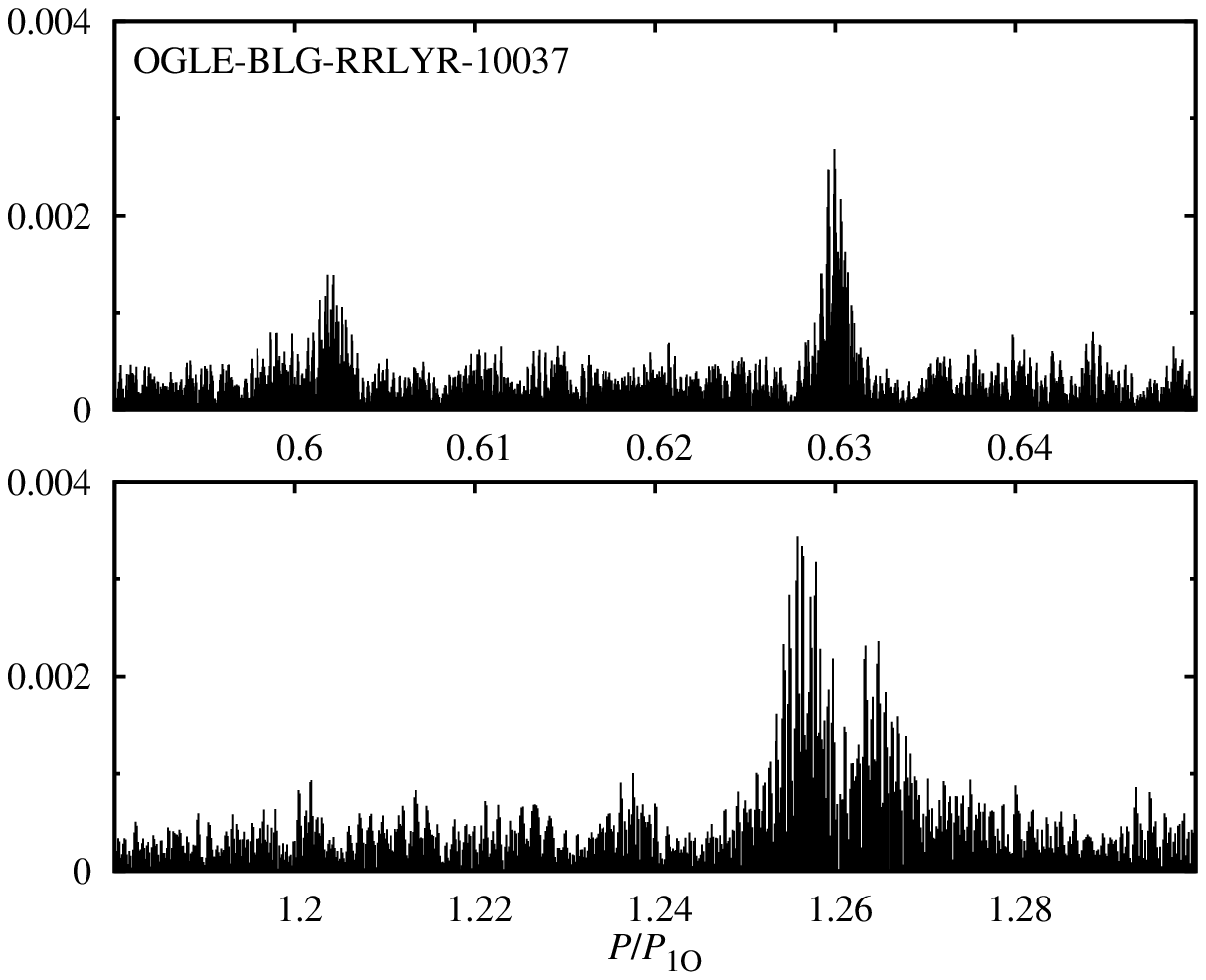}}
\resizebox{0.32\hsize}{!}{\includegraphics{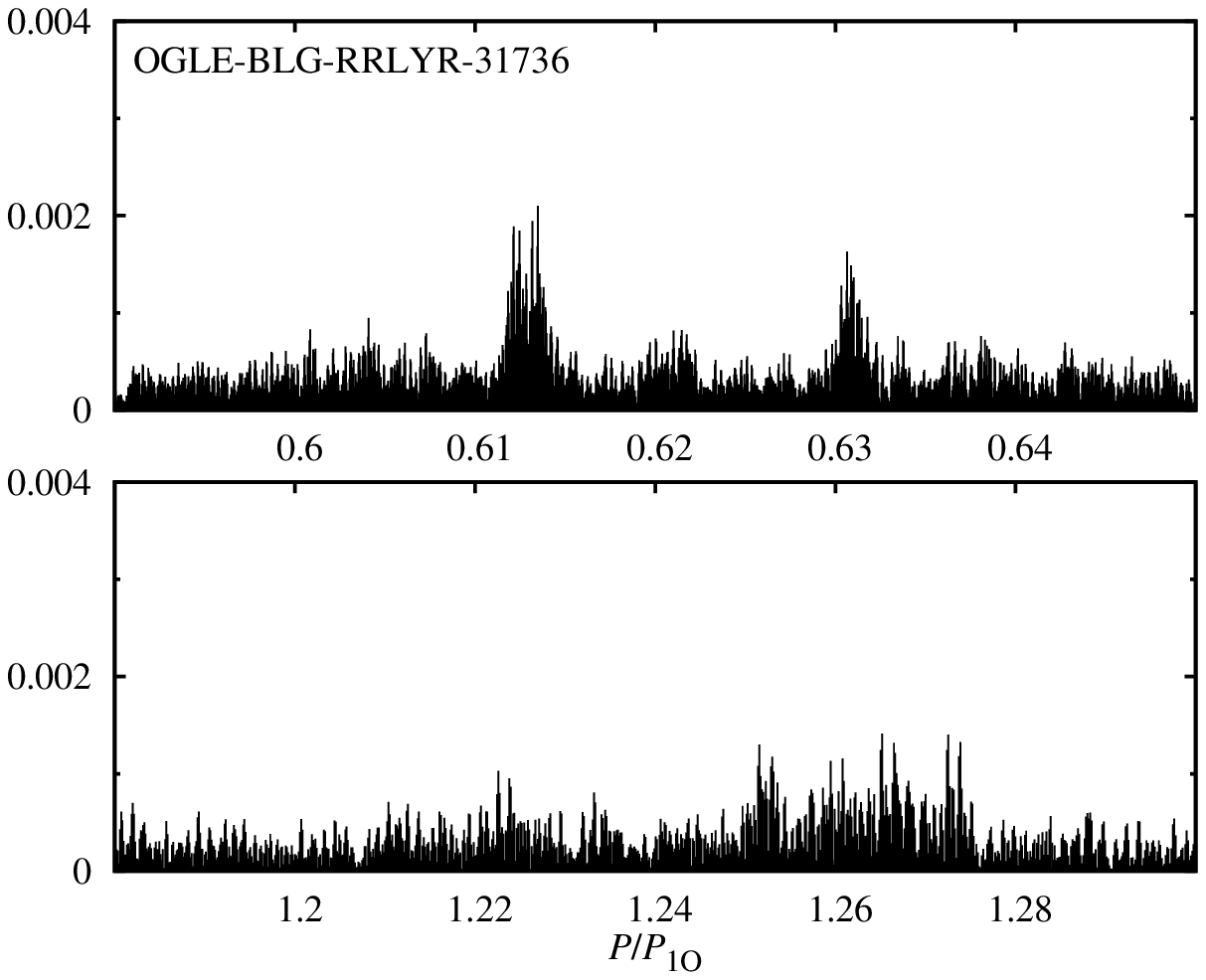}}
\caption{Frequency spectra centered at frequency range
         characteristic for additional mode (top panels) and its
         $1/2$ subharmonic (bottom panels) for a sample of 0.61
         stars. Horizontal axis scale is chosen such, that directly
         underneath a signal with frequency $f$ (top panel) is
         located its subharmonic with frequency $1/2f$ (the bottom
         panel).}
\label{fig.subharm}
\end{figure*}

A selection of 12 stars which show significant power excess at
subharmonic frequency range is presented in Fig.~\ref{fig.subharm}.
There are two panels for each star. Upper panel shows vicinity of
the additional mode(s), $\fx$. Lower panel shows vicinity of
its(their) subharmonic(s) at $1/2\fx$. The ranges of the two panels
are chosen in such a way, that subharmonic of each frequency in the
upper panel is located exactly underneath in the bottom panel.

Signals at subharmonic frequency range show more diverse structures
than detected at the parent frequency range. We find wide bands of
power excess. Again, it is difficult to select one peak to
characterize a band. Typically, a signal at subharmonic frequency range
fits well to a cluster of peaks detected at the parent frequency
range. Examples are OGLE-BLG-RRLYR-08826, OGLE-BLG-RRLYR-07448 in
Fig.~\ref{fig.subharm}. In general, bands of power at subharmonic
frequency range are wider than clusters of peaks detected at a
parent frequency range, while amplitudes of the highest peaks are
often comparable or even higher at the subharmonic range.

Very interesting stars are OGLE-BLG-RRLYR-08597, -06617, -10037,
-05071. In these stars there are clusters of peaks at $\fx$ and
$1/2\fx$, but the bands of power at subharmonic frequency range
seems to be split into two subbands centered at $\sim 1/2\fx$.
Exactly at $1/2\fx$ the signal is suppressed. In several stars the
bands of power excess are very wide. Examples of such stars in
Fig.~\ref{fig.subharm} are: OGLE-BLG-RRLYR-31736, -06461, -07486.

Other interesting stars are those that simultaneously show
significant signals at frequencies corresponding to the three
sequences, 0.61, 0.62 and 0.63. Not all of these signals have
counterparts in the subharmonic frequency range.
OGLE-BLG-RRLYR-08826 presented in Fig.~\ref{fig.subharm} is a good
example. In this star there are three clusters of peaks from the
three sequences, but only a subharmonic of $\sim0.63$ cluster is
present. OGLE-BLG-RRLYR-06610 is another good example illustrated in
Fig.~\ref{fig.subharm}.

In a frequency spectrum of OGLE-BLG-RRLYR-05296 the dominant signal
corresponds to $0.62$ sequence. However, in the subharmonic
frequency range a significant signal appears at frequency expected
for the 0.63 sequence. Similar situation is visible in
OGLE-BLG-RRLYR-08460. In the parent frequency range of this star we
detect peaks corresponding to $\sim 0.63$ sequence, but in the
subharmonic frequency range a signal appears at frequency expected
for the $\sim 0.61$ sequence. This may be a consequence of strong
time dependence of the observed phenomena, already discussed in the
previous Section. 

\subsection{Blazhko effect}\label{ssec:blazhko}

In spectra of some stars from the input sample (485 stars) we
detected equidistant triplets and doublets at the frequency of the
first overtone and its harmonics. These are signatures of the
Blazhko effect \citep{szabo14}. Such stars constitute $\approx\!10$\thinspace
per cent of the sample. Precise statistics and analysis of the
Blazhko effect in the Galactic bulge RRc stars will be a subject of
a forthcoming publication. We examined these stars for the presence
of the additional mode and found it in two of them (`bl' in the
remarks column of Tab.~\ref{tab.list}). We note that irregular phase
(and also amplitude) changes are frequent in RRc stars. This
non-stationarity is manifested in the frequency spectrum as a
residual signal remaining close to first overtone frequency after
prewhitening. This signal is either unresolved from the first
overtone frequency, or only marginally resolved. All those
stars are marked with `a' in the remarks column of
Tab.~\ref{tab.list}.

\begin{figure}
\centering
\resizebox{\hsize}{!}{\includegraphics{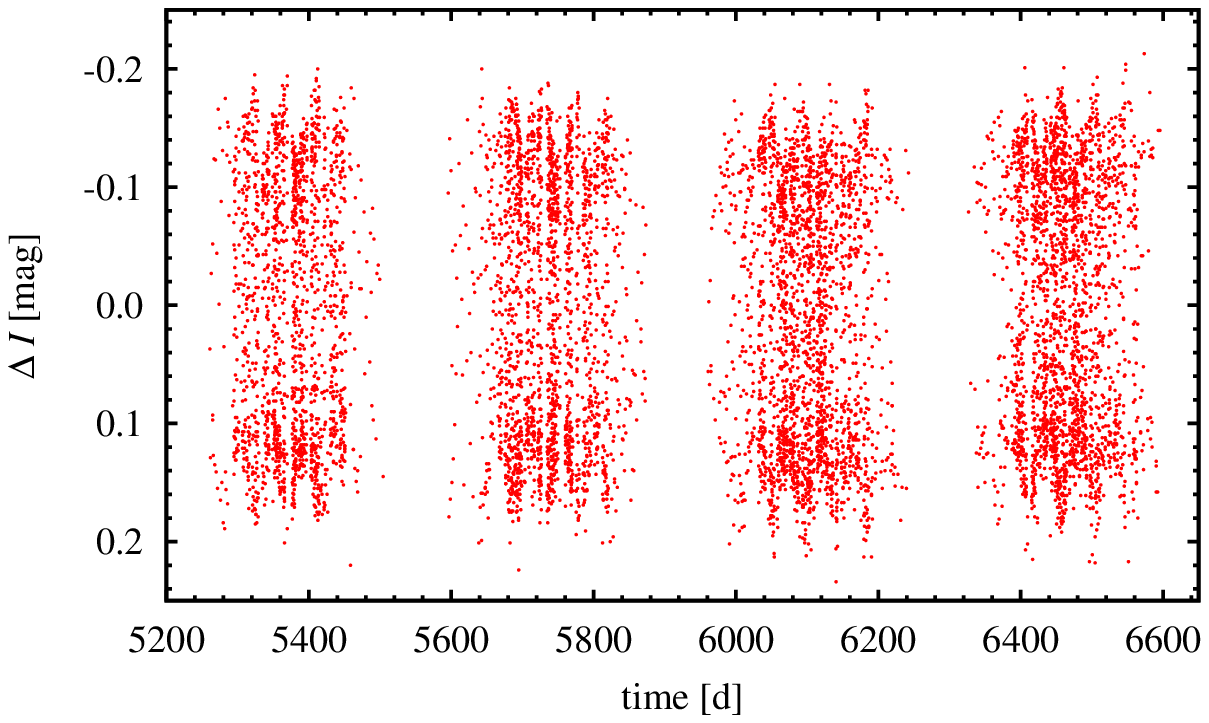}}
\caption{Light curve for OGLE-BLG-RRLYR-08177. Blazhko effect is clearly visible.}
\label{fig.blazko-data}
\end{figure}

The first star showing simultaneously the additional non-radial mode
and the Blazhko effect is OGLE-BLG-RRLYR-08177. Full light curve
solution for this star is provided in Tab.~\ref{tab:bl1}. The
additional non-radial mode, detected with $S/N=5.4$, fits the 0.61
sequence in the Petersen diagram, as $\px/\po=0.614$. Blazhko effect
is obvious in this star. Modulation is visible even in the raw data
which we display in Fig.~\ref{fig.blazko-data}. In the frequency
spectrum equidistant triplets are well visible; separation between
triplet components corresponds to modulation period $45.394 \pm
0.003$\thinspace days. At second harmonic a quintuplet is detected.
A signal at modulation frequency, $f_{\rm BL}$, is also detected in
the spectrum.

The second star with additional non-radial mode and Blazhko effect
is OGLE-BLG-RRLYR-32252. Full light curve solution for this star is
provided in Tab.~\ref{tab:bl2}. In this case, the Blazhko modulation
has much longer period: $494.3\pm2.6$ days. As length of the OGLE-IV
data is about 1336 days, the triplet components are resolved, but
weakly. Unfortunately, there is no OGLE-III data for this star. In
the frequency spectrum triplets are detected at $\fo$, $3\fo$, and
at $4\fo$ only higher frequency side-peak is detected. The star
shows the additional non-radial mode that fits the 0.63 sequence.
Frequency combination, $\fx+\fo$, is detected in the spectrum as
well.

\subsection{Comparison with other studies}\label{ssec:comp}

\begin{figure}
\centering
\resizebox{\hsize}{!}{\includegraphics{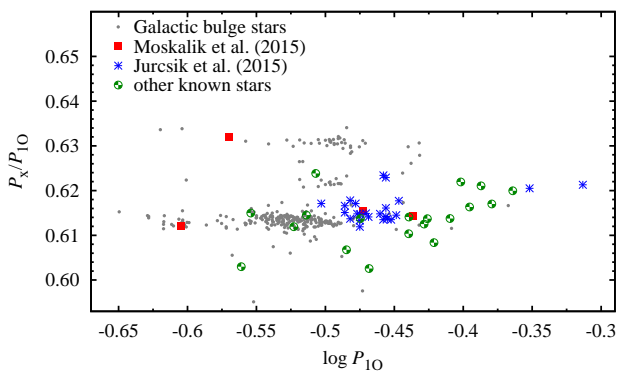}}
\caption{
Petersen diagram for all known 0.61 stars. The Galactic bulge sample is plotted with gray dots. {\it Kepler} stars \citep{pamsm15} and M3 stars \citep{jurcsik_M3} are plotted with red and blue symbols, respectively. Green symbols are used for other known 0.61 stars \citep[data from tab.~8 of][see references therein]{pamsm15}.}
\label{fig.pet2}
\end{figure}

The subject of this paper, stars with additional non-radial mode
with $\px/\po\approx 0.61$, were also detected in other stellar
populations, not only in the Galactic bulge. 
All known stars with additional non-radial mode are plotted in the 
Petersen diagram in Fig.~\ref{fig.pet2}. In this Section we
compare our results with two studies. First is a study of
\cite{pamsm15} who analysed {\it Kepler} observations of four RRc
stars in the Cygnus field and summarized ground and space observations of other 
known 0.61 stars. As mentioned in the Introduction, 13 out of 14 RRc/RRd stars
observed from space show the additional mode (see also footnote 2 on
page 1). In the second study, \cite{jurcsik_M3} studied the first
overtone and double-mode RR~Lyr stars in the globular cluster M3.
In M3 the additional mode of interest was detected in 14 RRc stars
(out of 37 identified as RRc) and in 4 RRd stars (out of 10); the
fraction of stars showing the additional mode in M3 is thus
38\thinspace per cent for RRc and 40\thinspace per cent for RRd
stars. In our study 27\thinspace per cent of RRc stars shows the
additional mode (analysed sample of 4 RRd stars, of which none shows
the additional mode, is too small to draw any conclusions).

In three stars observed by {\it Kepler} period ratios of additional
mode to first overtone are 0.612, 0.614 and 0.616. Hence, these
stars well fit the 0.61 sequence on the Petersen diagram
(Fig.~\ref{fig.pet2}). One star has period ratio 0.632 and fits the
0.63 sequence. Most of the stars from M3 have period ratios placing
them in the 0.61 sequence and in four stars period ratios correspond
to 0.62 sequence. Thus, the three sequences revealed in the
observations of the Galactic bulge stars seem to be present in stars
from other populations as well. Very interesting result of the
present study is that in single star three or two additional modes,
each corresponding to one of the three sequences, may be visible
simultaneously. This phenomenon is present in one star from M3,
which shows two additional modes, of the 0.61 and 0.62 sequences.
We also note that population effect is clearly visible in Fig.~\ref{fig.pet2}, when comparing the two most homogeneous samples, the Galactic bulge and M3 samples. The M3 sample covers a different period range than the Galactic bulge sample. It is also clear that the lowest sequence has, on average, a slightly larger period ratio in the M3 sample. All other stars are scattered over the diagram. Population effect is clear, but what causes the differences remains unknown. Metallicity is a first guess. We know that metallicity is an important factor influencing the period ratios of the pulsation modes. Surprisingly, the stars from M3, where the metallicity is homogeneous \citep{clementini}, show larger scatter in the Petersen diagram, than the Galactic bulge sample, where the metallicity spread is significant \citep{smolec2005}. What causes the difference between the properties of M3 and Galactic bulge populations remains unclear. 

We note that additional non-radial modes, with characteristic period
ratio, $\px/\po\in(0.61,\, 0.64)$, are also detected in classical
Cepheids in both Magellanic Clouds and in one star in the Galactic
disc \citep{mk09,ogle_cep_lmc,ogle_cep_smc,pietruk}. In the Petersen
diagram \citep[see fig.~2 in][]{pam14}, these stars also form three nearly parallel and well separated 
sequences, with clear trend of decreasing period ratio
with increasing period. All sequences are well populated, with largest 
number of stars in the lowest sequence. In the case of RR~Lyr stars, 
majority of stars fall into the lowest sequence and spacing between 
sequences in the Petersen diagram is smaller.

Amplitude of the additional mode is always much lower than the
amplitude of the first overtone. In stars of the Galactic bulge the
amplitude ratio ranges from 0.6 to 5.5\thinspace per cent, in {\it
Kepler} stars it is in a range of 2 to 5\thinspace per cent, and in
M3 stars in a range of 2 to 8\thinspace per cent.

Properties of additional mode seem very similar in all known stars,
especially its non-stationarity is visible in most of them.
Variations of amplitudes and phases are very irregular and could be
studied best with nearly continuous {\it Kepler} data.
Time-dependent Fourier analysis reported in \cite{pamsm15} shows,
that time scale of the variability ranges from 10\thinspace d to
200\thinspace d. As a result of this variability, in the frequency
spectra of these stars we do not detect a single, coherent secondary
peak, but either a broadened peak, or a multiplet of approximately
equally spaced broadened peaks \citep[see figs. 8 and 12
in][]{pamsm15}. Similar structures are also present in the frequency
spectra of stars from the Galactic bulge (see Fig.~\ref{fig.struct} 
for some examples). In
particular, in the frequency spectra of several stars from our
sample, at the frequency of additional mode, we detect triplets and
doublets. If these structures result from-quasi-periodic modulation
of additional mode, as in case of {\it Kepler} stars, we can
estimate the time scales of this modulation from separation of the
peaks. It ranges from $\approx\!20$\thinspace d to
$\approx\!120$\thinspace d, with typical values corresponding to
$40-60$\thinspace d.

\begin{figure}
\centering
\resizebox{\hsize}{!}{\includegraphics{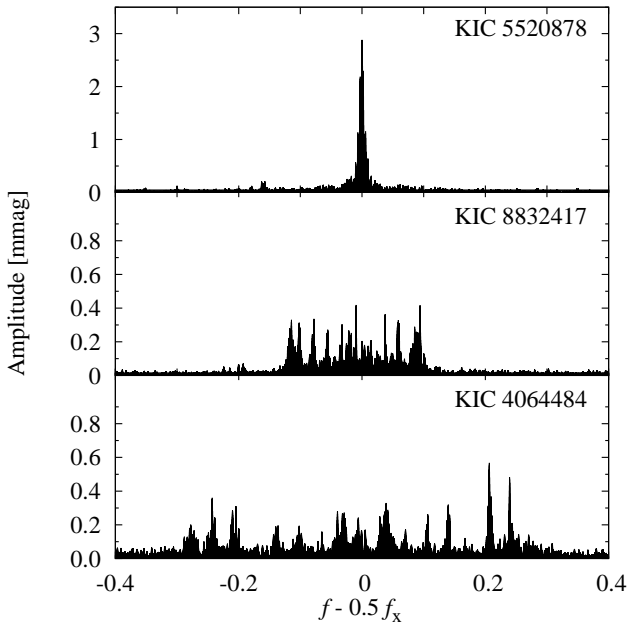}}
\caption{Frequency spectra of {\it Kepler} stars centered at subharmonic frequency $0.5 \fx$.}
\label{fig.kep05}
\end{figure}

In all four stars observed by {\it Kepler}, significant signals at
subharmonics of the additional mode, both at $1/2\fx$ (in 3 stars)
and at $3/2\fx$ (in 4 stars) were detected. In OGLE-III photometry
we found subharmonics at $1/2\fx$ in four stars only. Much better
OGLE-IV data allowed to find subharmonic frequency in 20\thinspace
per cent of 0.61 stars, interestingly, nearly exclusively at around
$1/2\fx$. As discussed in Sect.~\ref{ssec:subh} and well visible in
Fig.~\ref{fig.subharm} signals at $1/2\fx$ have complex structure
and appear as wide bands of power excess rather than sharp peaks
exactly at $1/2\fx$. It is also the case for {\it Kepler} stars. The
structure of sub-harmonics at $3/2\fx$ is plotted in fig.~8 of
\cite{pamsm15}. In Fig.~\ref{fig.kep05} we plot the frequency
spectra centered at $1/2\fx$ for the three {\it Kepler} stars that
show power excess here. Only in KIC5520878 a relatively sharp peak,
centered at $1/2f_x$ is present. For two other stars very broad
bands of power excess at the vicinity of $1/2\fx$ frequency are
present, with several resolved peaks of about the same height. These
structures are similar to that observed in the Galactic bulge stars
(Fig.~\ref{fig.subharm}). Interestingly, amplitude of highest peak
at around $1/2\fx$ in the Galactic bulge stars might be higher than
amplitudes of the additional mode. Examples include
OGLE-BLG-RRLYR-08597, -10037, -05071, -08460, -07448 in
Fig.~\ref{fig.subharm}. Subharmonics were not detected in M3 stars.

\section{Summary and conclusions}\label{sec:summary}

We analyzed OGLE-IV photometry of the first overtone and double-mode
RR Lyrae stars (RRc and RRd stars, respectively) of the Galactic
bulge. We limited the scope of the study to two fields with highest
cadence, for which more than 8000 measurements have been accumulated
over 4 years. Our sample consists of 485 RRc stars and 4 RRd stars.
The main goal of this project was to search for secondary non-radial
modes with period ratio to the first radial overtone of $P_{\rm
x}/P_{\rm 1O} \sim 0.61$.

The most important results of our study can be summarized as
follows.

\begin{itemize}
\item Low amplitude secondary modes, $\fx$, with period ratio of $\sim
      0.61$ are detected in 131 RRc stars. This constitutes 27 per
      cent of the studied sample. The inferred occurrence rate is an
      order of magnitude higher than determined for the Galactic
      bulge with the OGLE-III data \citep[3 per cent,][]{netzel} and
      is comparable to that found in the globular cluster M3
      \citep[38 per cent,][]{jurcsik_M3}. No additional modes beyond
      the radial ones are detected in the 4 RRd stars of our sample.

\item The amplitudes of the secondary modes are very low, in the mmag
      range and do not exceed 5.5 per cent of the amplitude of the
      dominant first radial overtone. This upper limit is somewhat
      lower, but similar to that found in other studies
      \citep[e.g.][]{om09, pamsm15, netzel, jurcsik_M3, molnar}.

\item We detect the secondary modes in two stars displaying a long-term
      modulation of the radial mode (Blazhko effect). Three similar
      variables have also been found recently in M3
      \citep{jurcsik_M3}. Thus, excitation of non-radial mode with
      $P_{\rm x}/P_{\rm 1O} \sim 0.61$ and the Blazhko modulation of
      the radial mode are not mutually exclusive.

\item Combining the results of this paper with those of our previous
      analysis of OGLE-III data \citep{netzel}, we now know 262
      Galactic bulge RR~Lyrae stars (260 RRc and 2 RRd), in which
      non-radial modes with period ratio of $\sim 0.61$ are excited.
      This is by far the largest homogeneous sample of this type of
      objects.

\item The period ratios measured in Galactic bulge RRc/RRd stars are
      all in a narrow range of $P_{\rm x}/P_{\rm 1O} = 0.598-0.634$.
      When plotted on the period ratio vs. period diagram (so-called
      Petersen diagram), these stars form three, approximately
      parallel sequences. The sequences are nearly horizontal and
      are centered at $P_{\rm x}/P_{\rm 1O} \approx\!0.613$, $\approx\!0.623$ 
      and $\approx\!0.631$. The lowest sequence is by far most
      populated. The sequence at $\approx\!0.631$ was tentatively
      identified by \cite{netzel} and is now fully confirmed. The
      middle sequence at $\approx\!0.623$ is identified for the first
      time.

\item The existence of the three sequences of RRc stars is analogous to
      behaviour of the first overtone Cepheids. In the later case,
      non-radial modes with $P_{\rm x}/P_{\rm 1O} = 0.60-0.65$ have
      been detected in more than 170 objects. These Cepheids form
      three parallel sequences in the Petersen diagram as well
      \citep[for summary see][]{pam14}.

\item In 20 OGLE-IV RRc stars we detected more than one secondary mode.
      In 14 variables two secondary modes belonging to two different
      sequences are present. In 6 variables three secondary modes
      representing all three sequences are present. Clearly, modes
      of different sequences can be excited simultaneously in the
      same star.

\item Secondary modes are in many cases non-stationary, changing their
      amplitude and frequency from season to season and during the
      seasons. Consequently, in the Fourier transform of entire
      dataset, they do not appear as single, coherent peaks, but
      rather as broadened peaks or as narrow bands of power
      (clusters of peaks).

\item In 7 OGLE-IV RRc stars a secondary mode is split into three
      almost equally spaced clusters of peaks. In further 10 stars a
      secondary mode is split into two components. As was shown in
      case of {\it Kepler} RRc stars, such structure might result
      from quasi-periodic modulation of the mode \citep[][see their
      Fig.\thinspace 12]{pamsm15}. The splittings observed in
      OGLE-IV RRc stars correspond to modulation timescales of
      $20-120$\thinspace d. Both non-stationarity and splitting
      occur most frequently for modes of sequence 0.61.

\item In 26 OGLE-IV RRc stars a significant signal at around $\sim
      1/2\fx$, i.e. at around a subharmonic of the secondary
      frequency is found. This constitutes 20 per cent of all 0.61
      stars identified in this study. A subharmonic at around $\sim
      3/2\fx$ is found in two stars. Until now, subharmonics of
      $\fx$ have been detected only in RRc and RRd variables
      observed from space. With the OGLE data, we have for the first
      time a possibility to see them from the ground (see also
      Netzel et al. 2015). Judging from space photometry,
      subharmonics are common and occur in at least 75 per cent of
      the 0.61 variables \citep{pamsm15}. Their presence is a
      signature of a period doubling of the secondary oscillation
      \citep[e.g.][]{blherPD}.

\item Subharmonic signal is highly non-coherent. We always detect a
      band of power, located at around the subharmonic frequency,
      $1/2\fx$. This band of power is usually broader and sometimes
      much broader than the cluster of peaks corresponding to the
      parent mode, $\fx$. In addition, it often displays a wide,
      essentially flat structure or a bimodal structure, not seen in
      the parent mode. When more than one secondary mode is present, subharmonics structures often are seen only for one of them, and this is not always the one with the largest amplitude. We note, that similar broad bands of power at
      $1/2\fx$ are also detected in two RRc stars studied by {\it
      Kepler} space telescope \citep[][and our
      Fig.~\ref{fig.kep05}]{pamsm15}. Clearly, the subharmonics of
      the secondary modes are non-stationary to even larger degree
      than the modes themselves.
\end{itemize}

The discussed stars with additional non-radial mode are not well understood. We do not know which non-radial mode is excited and what is the mechanism behind its excitation. The only and unsuccessful attempt to understand these stars is a work by \cite{wd12} focused on classical Cepheids. Recently, \cite{lindner} noticed that period ratio in the discussed stars is close to the reciprocal of the golden ratio, which is $\approx\!0.618033$, and argued that pulsation with such a period ratio of the excited modes should be stable against perturbations. They refer to 0.61 stars as {\it golden stars}. A glimpse at the Petersen diagram (Fig.~\ref{fig.pet2}) shows, that such explanation is unlikely. The stars are indeed located close to the reciprocal of the golden ratio, but do not cluster at it. The reciprocal of the golden ratio falls in between the lowest and the middle sequences and clearly {\it does not attract} any of the stars. The proximity of the period ratio in 0.61 stars to the reciprocal of the golden ratio is a pure coincidence in our opinion, and explanation for the stars, on solid grounds of pulsation theory, must be searched for.

%
%
%
%

\section*{Acknowledgements}
This research is supported by the Polish National Science Centre through grant DEC-2012/05/B/ST9/03932.

We are greateful to the referee, Johanna Jurcsik, for constructive comments, which helped to improve this publication.





\begin{thebibliography}{99}
\bibitem[\protect\citeauthoryear{Alcock et al.}{2003}]{alcock} Alcock C., Alves D.R., Becker Al. et al., 2003, ApJ, 598, 597
\bibitem[\protect\citeauthoryear{Clementini et al.}{2004}]{clementini} Clementini G., Corwin. T.M., Carney B.W., Sumerel A.N., 2004, ApJ, 127, 938
\bibitem[\protect\citeauthoryear{Dziembowski}{2012}]{wd12} Dziembowski W., 2012, Acta Astron., 62, 323
\bibitem[\protect\citeauthoryear{Gruberbauer et al.}{2007}]{aqleo} Gruberbauer M., Kolenberg K., Rowe J. et al., 2007, MNRAS, 379, 1498
\bibitem[\protect\citeauthoryear{Jurcsik et al.}{2009}]{jurcsik_KBS} Jurcsik J., et al., 2009, MNRAS, 400, 1006
\bibitem[\protect\citeauthoryear{Jurcsik et al.}{2014}]{jurcsik_BLRRd} Jurcsik J., Smitola P., Hajdu G., Nuspl J., 2014, ApJ, 797, L3
\bibitem[\protect\citeauthoryear{Jurcsik et al.}{2015}]{jurcsik_M3} Jurcsik J., et al., 2015, ApJ Suppl. Ser., in press, arXiv:1504.06215
\bibitem[\protect\citeauthoryear{Kolenberg et al.}{2010}]{kol10} Kolenberg K., et al., 2010, ApJ, 713, L198
\bibitem[\protect\citeauthoryear{Lindner et al.}{2015}]{lindner} Lindner J.F., Kohar V., Kia B., Hippke M., Learned J.G., Ditto, W.L., 2015, Phys. Rev. Lett., 114, 054101
\bibitem[\protect\citeauthoryear{Moln\'ar et al.}{2015}]{molnar} Moln\'ar L., et al. 2015, MNRAS, submitted
\bibitem[\protect\citeauthoryear{Moskalik}{2013}]{pam13} Moskalik P., 2013, in Su\'arez J.C., Garrido R., Balona L.A., Christensen-Dalsgaard J., eds, Astrophysics and Space Sci. Proc. 31, Stellar Pulsations: Impact of New Instrumentation and New Insights. Springer-Verlag, Berlin, Heidelberg, p. 103
\bibitem[\protect\citeauthoryear{Moskalik}{2014}]{pam14} Moskalik P., 2014, in Guzik J.A., Chaplin W.J., Handler G., Pigulski A., eds, IAU~Symp.~301, p.~249
\bibitem[\protect\citeauthoryear{Moskalik \& Ko\l{}aczkowski}{2009}]{mk09} Moskalik P., Ko\l{}aczkowski Z., 2009, MNRAS, 394, 1649
\bibitem[\protect\citeauthoryear{Moskalik et al.}{2013}]{pam+13} Moskalik P., et al., 2013, in Su\'arez J.C., Garrido R., Balona L.A., Christensen-Dalsgaard J., eds, Astrophysics and Space Sci. Proc. 31, Stellar Pulsations: Impact of New Instrumentation and New Insights. Springer-Verlag, Berlin, Heidelberg, poster no. 34; Online data at http://dx.doi.org/10.1007/978-3-642-29630-7\_53
\bibitem[\protect\citeauthoryear{Moskalik et al.}{2015}]{pamsm15} Moskalik P., Smolec R., Kolenberg K. et al., 2015, MNRAS, 447, 2348
\bibitem[\protect\citeauthoryear{Netzel, Smolec \& Moskalik}{2015}]{netzel} Netzel H., Smolec R., Moskalik P., 2015, MNRAS, 447, 1173
\bibitem[\protect\citeauthoryear{Netzel, Smolec \& Dziembowski}{2015}]{netzel68} Netzel H., Smolec R., Dziembowski W., 2015, MNRAS Lett., 451, L25
\bibitem[\protect\citeauthoryear{Olech \& Moskalik}{2009}]{om09} Olech A., Moskalik P., 2009, A\&A, 494, L17
\bibitem[\protect\citeauthoryear{Pietrukowicz et al.}{2013}]{pietruk} Pietrukowicz P., et al., 2013, Acta Astron., 63, 379
\bibitem[\protect\citeauthoryear{Smolec}{2005}]{smolec2005} Smolec R., 2005, Acta Astron., 55, 59
\bibitem[\protect\citeauthoryear{Smolec et al.}{2012}]{blherPD} Smolec R., Soszy\'nski I., Moskalik P., et al., 2012, MNRAS, 419, 2407
\bibitem[\protect\citeauthoryear{Smolec et al.}{2015}]{rs15a} Smolec R., Soszy\'nski I., Udalski A. et al., 2015, MNRAS, 447, 3756
\bibitem[\protect\citeauthoryear{Soszy\'nski et al.}{2008}]{ogle_cep_lmc} Soszy\'nski I., Poleski R., Udalski A., et al., 2008, Acta Astron., 58, 163
\bibitem[\protect\citeauthoryear{Soszy\'nski et al.}{2009}]{ogle_rr_lmc} Soszy\'nski I., Udalski A., Szyma\'nski M.K. et al., 2009, Acta Astron., 59, 1
\bibitem[\protect\citeauthoryear{Soszy\'nski et al.}{2010}]{ogle_cep_smc} Soszy\'nski I., Poleski R., Udalski A., et al., 2010, Acta Astron., 60, 17
\bibitem[\protect\citeauthoryear{Soszy\'nski et al.}{2011}]{ogle_rr_blg} Soszy\'nski I., Dziembowski W., Udalski A., et al., 2011, Acta Astron., 61, 1
\bibitem[\protect\citeauthoryear{Soszy\'nski et al.}{2014}]{ogleiv} Soszy\'nski I., Udalski A., Szyma\'nski M.K. et al., 2014, Acta Astron., 64, 177
\bibitem[\protect\citeauthoryear{Szab\'o}{2014}]{szabo14} Szab\'o R., 2014, IAUS, 301, 241
\bibitem[\protect\citeauthoryear{Szab\'o et al.}{2010}]{szabo10} Szab\'o R., Koll\'ath Z., Moln\'ar L. et al., 2010, MNRAS, 409, 1244
\bibitem[\protect\citeauthoryear{Szab\'o et al.}{2014}]{szabo_corot} Szab\'o R., Benk\H{o} J.M., Papar\'o M., 2014, A\&A, 570, A100
\bibitem[\protect\citeauthoryear{S\"uveges et al.}{2012}]{sdss} S\"uveges M., Sesar B., V\'aradi M. et al., 2012, MNRAS, 424, 2528
\bibitem[\protect\citeauthoryear{Udalski et al.}{2008}]{ogleIII} Udalski A., Szyma\'nski M.K., Soszy\'nski I., Poleski R., 2008, Acta Astron., 58, 69
\bibitem[\protect\citeauthoryear{Udalski, Szyma\'nski \& Szyma\'nski}{2015}]{ogleiv_tech} Udalski A., Szyma\'nski M.K., Szyma\'nski G., 2015, Acta Astron., 65, 1

\end{thebibliography}




\appendix

\section[]{List of 0.61 stars}
\begin{table*} 
 \centering
 \begin{minipage}{140mm}
  \caption{Properties of stars with non-radial mode (OGLE-IV)}
  \label{tab.list}
  
  \begin{tabular}{@{}lccccccl@{}}
  \hline
  \multicolumn{8}{l}{a - period change; bl - Blazhko effect; td - signal visible only after time-dependent prewhitening;}\\ 
  \multicolumn{8}{l}{c - combination frequency; s - power excess at subharmonic; d - dublet; t - triplet; e - additional frequency;}\\
  \multicolumn{8}{l}{g - non-stationary $\fx$; f - complex structure of $\fx$; sr - half of the first season removed}\\ 
  \multicolumn{8}{l}{h - stars detected in OGLE-III data \citep{netzel}}\\ 
 \hline
 \hline
Name & $\po$\thinspace[d] & $\px$\thinspace[d] & $\px/\po$ & $\ao$\thinspace[mag] & $\ax$\thinspace[mag] & $\ax/\ao$ &  Remarks            \\
 \hline
OGLE-BLG-RRLYR-04067 & 0.31994 & 0.19592 & 0.61236 & 0.10687 & 0.00479 & 0.0448 & a \\
OGLE-BLG-RRLYR-04105 & 0.30582 & 0.18724 & 0.61226 & 0.13009 & 0.00476 & 0.0366 &  \\
OGLE-BLG-RRLYR-04549 & 0.29964 & 0.18874 & 0.62990 & 0.12246 & 0.00384 & 0.0313 &  \\
OGLE-BLG-RRLYR-04599 & 0.28939 & 0.17797 & 0.61498 & 0.13980 & 0.00340 & 0.0243 & a,g \\
OGLE-BLG-RRLYR-04754 & 0.28631 & 0.17578 & 0.61394 & 0.12788 & 0.00453 & 0.0354 & g,h \\
OGLE-BLG-RRLYR-04762 & 0.29465 & 0.18061 & 0.61295 & 0.12640 & 0.00700 & 0.0554 &  \\
OGLE-BLG-RRLYR-04902 & 0.32219 & 0.19728 & 0.61230 & 0.12395 & 0.00314 & 0.0253 & g \\
OGLE-BLG-RRLYR-04942 & 0.31855 & 0.19534 & 0.61320 & 0.09610 & 0.00305 & 0.0317 & sr \\
OGLE-BLG-RRLYR-04974 & 0.29665 & 0.18163 & 0.61228 & 0.12452 & 0.00436 & 0.0350 & t \\
OGLE-BLG-RRLYR-04989 & 0.30670 & 0.18767 & 0.61191 & 0.13024 & 0.00247 & 0.0190 & a,g \\
OGLE-BLG-RRLYR-05071 & 0.33469 & 0.21078 & 0.62978 & 0.13613 & 0.00255 & 0.0187 & a,s \\
OGLE-BLG-RRLYR-05141 & 0.29274 & 0.17963 & 0.61362 & 0.12807 & 0.00434 & 0.0339 &  \\
OGLE-BLG-RRLYR-05201 & 0.29693 & 0.18220 & 0.61361 & 0.12856 & 0.00453 & 0.0353 & a,g \\
OGLE-BLG-RRLYR-05231 & 0.28601 & 0.17575 & 0.61450 & 0.14483 & 0.00178 & 0.0123 & a,g \\
OGLE-BLG-RRLYR-05266 & 0.30485 & 0.18925 & 0.62079 & 0.12610 & 0.00294 & 0.0233 & a,g \\
 & 0.30485 & 0.18670 & 0.61244 & 0.12610 & 0.00217 & 0.0172 &  \\
OGLE-BLG-RRLYR-05291 & 0.32072 & 0.19692 & 0.61398 & 0.11398 & 0.00229 & 0.0201 & a,c \\
OGLE-BLG-RRLYR-05296 & 0.31873 & 0.19823 & 0.62195 & 0.13114 & 0.00291 & 0.0222 & c,g,s \\
OGLE-BLG-RRLYR-05301 & 0.30562 & 0.18729 & 0.61282 & 0.12450 & 0.00554 & 0.0445 & a,c,g,h \\
OGLE-BLG-RRLYR-05527 & 0.28715 & 0.17593 & 0.61266 & 0.12418 & 0.00230 & 0.0185 & c,f \\
OGLE-BLG-RRLYR-05531 & 0.30808 & 0.18908 & 0.61372 & 0.09446 & 0.00304 & 0.0322 & g \\
OGLE-BLG-RRLYR-05542 & 0.28751 & 0.17730 & 0.61668 & 0.14280 & 0.00154 & 0.0108 & a,f \\
OGLE-BLG-RRLYR-05600 & 0.30881 & 0.18947 & 0.61355 & 0.13017 & 0.00340 & 0.0261 & c,t \\
OGLE-BLG-RRLYR-05672 & 0.28435 & 0.17464 & 0.61420 & 0.14864 & 0.00170 & 0.0114 & g \\
OGLE-BLG-RRLYR-05898 & 0.32687 & 0.20613 & 0.63060 & 0.13245 & 0.00361 & 0.0272 & a \\
OGLE-BLG-RRLYR-05924 & 0.29997 & 0.18412 & 0.61380 & 0.14345 & 0.00447 & 0.0312 & a,c \\
OGLE-BLG-RRLYR-05928 & 0.31043 & 0.19071 & 0.61434 & 0.10101 & 0.00129 & 0.0128 & a \\
OGLE-BLG-RRLYR-05934 & 0.29263 & 0.17948 & 0.61334 & 0.13069 & 0.00279 & 0.0214 & a,g \\
OGLE-BLG-RRLYR-05937 & 0.31421 & 0.19241 & 0.61235 & 0.12346 & 0.00370 & 0.0300 & a,g \\
OGLE-BLG-RRLYR-05965 & 0.31396 & 0.19266 & 0.61364 & 0.11521 & 0.00308 & 0.0267 & a,g \\
OGLE-BLG-RRLYR-06056 & 0.33646 & 0.20105 & 0.59756 & 0.15306 & 0.00179 & 0.0117 & a \\
OGLE-BLG-RRLYR-06083 & 0.30647 & 0.18756 & 0.61199 & 0.13596 & 0.00338 & 0.0249 & f \\
OGLE-BLG-RRLYR-06130 & 0.30384 & 0.18632 & 0.61323 & 0.09404 & 0.00222 & 0.0237 & d,e \\
OGLE-BLG-RRLYR-06143 & 0.30875 & 0.18947 & 0.61368 & 0.12086 & 0.00282 & 0.0233 & c,g \\
OGLE-BLG-RRLYR-06149 & 0.27854 & 0.17084 & 0.61334 & 0.10532 & 0.00194 & 0.0184 & d \\
OGLE-BLG-RRLYR-06194 & 0.30451 & 0.19077 & 0.62650 & 0.11891 & 0.00168 & 0.0141 & a \\
OGLE-BLG-RRLYR-06200 & 0.28699 & 0.17647 & 0.61490 & 0.14828 & 0.00306 & 0.0206 & a,c,t \\
OGLE-BLG-RRLYR-06265 & 0.32974 & 0.20719 & 0.62833 & 0.12563 & 0.00223 & 0.0178 & a \\
OGLE-BLG-RRLYR-06420 & 0.31401 & 0.19246 & 0.61290 & 0.10435 & 0.00348 & 0.0333 & a,c,h \\
 & 0.31401 & 0.19542 & 0.62233 & 0.10435 & 0.00188 & 0.0181 &  \\
OGLE-BLG-RRLYR-06461 & 0.29651 & 0.18719 & 0.63130 & 0.13055 & 0.00186 & 0.0142 &  \\
OGLE-BLG-RRLYR-06497 & 0.28119 & 0.17230 & 0.61276 & 0.15406 & 0.00524 & 0.0340 & c,e,t \\
OGLE-BLG-RRLYR-06571 & 0.29408 & 0.18155 & 0.61734 & 0.11915 & 0.00147 & 0.0124 & a \\
OGLE-BLG-RRLYR-06590 & 0.33661 & 0.21246 & 0.63118 & 0.13126 & 0.00228 & 0.0174 & a,e \\
OGLE-BLG-RRLYR-06610 & 0.31623 & 0.19649 & 0.62135 & 0.11623 & 0.00196 & 0.0169 & a,s \\
 & 0.31623 & 0.19377 & 0.61274 & 0.11623 & 0.00245 & 0.0211 &  \\
 & 0.31623 & 0.19963 & 0.63126 & 0.11623 & 0.00188 & 0.0162 &  \\
OGLE-BLG-RRLYR-06617 & 0.32632 & 0.20603 & 0.63137 & 0.13774 & 0.00251 & 0.0182 & a,g \\
OGLE-BLG-RRLYR-06627 & 0.29820 & 0.18072 & 0.60602 & 0.11553 & 0.00202 & 0.0175 & a,c,d,g \\
OGLE-BLG-RRLYR-06659 & 0.28984 & 0.17607 & 0.60746 & 0.11661 & 0.00222 & 0.0191 & a,c,d,g \\
OGLE-BLG-RRLYR-06693 & 0.29235 & 0.17935 & 0.61349 & 0.14231 & 0.00360 & 0.0253 & a,g \\
OGLE-BLG-RRLYR-06802 & 0.31830 & 0.19501 & 0.61267 & 0.11096 & 0.00213 & 0.0192 & a,c,g,s,td \\
 & 0.31830 & 0.19792 & 0.62181 & 0.11096 & 0.00207 & 0.0186 &  \\
 & 0.31830 & 0.20098 & 0.63142 & 0.11096 & 0.00169 & 0.0152 &  \\
OGLE-BLG-RRLYR-06885 & 0.28525 & 0.17527 & 0.61445 & 0.15038 & 0.00227 & 0.0151 & a,g \\
OGLE-BLG-RRLYR-07076 & 0.29638 & 0.18209 & 0.61437 & 0.12542 & 0.00202 & 0.0161 & a,t,h \\
\hline
 \end{tabular}
\end{minipage}
\end{table*}

\begin{table*} 
 \centering
  \begin{minipage}{140mm}
  \begin{tabular}{@{}lccccccl@{}}
   \hline
Name & $\po$\thinspace[d] & $\px$\thinspace[d] & $\px/\po$ & $\ao$\thinspace[mag] & $\ax$\thinspace[mag] & $\ax/\ao$ & Remarks            \\
  \hline
OGLE-BLG-RRLYR-07091 & 0.27903 & 0.17095 & 0.61266 & 0.14490 & 0.00216 & 0.0149 & a,t \\
OGLE-BLG-RRLYR-07094 & 0.26797 & 0.16458 & 0.61418 & 0.10938 & 0.00210 & 0.0192 & a,g,h \\
OGLE-BLG-RRLYR-07096 & 0.31880 & 0.19548 & 0.61317 & 0.11715 & 0.00263 & 0.0225 & a,c,d,g \\
OGLE-BLG-RRLYR-07103 & 0.30453 & 0.18608 & 0.61102 & 0.12448 & 0.00305 & 0.0245 &  \\
OGLE-BLG-RRLYR-07135 & 0.35978 & 0.22448 & 0.62392 & 0.10472 & 0.00138 & 0.0132 & a,s \\
OGLE-BLG-RRLYR-07292 & 0.32493 & 0.20495 & 0.63075 & 0.13019 & 0.00235 & 0.0181 & a,g,s \\
OGLE-BLG-RRLYR-07303 & 0.32197 & 0.20048 & 0.62267 & 0.12339 & 0.00193 & 0.0157 & a,c \\
OGLE-BLG-RRLYR-07375 & 0.32464 & 0.20176 & 0.62148 & 0.12227 & 0.00266 & 0.0218 & a,c,s \\
 & 0.32464 & 0.19891 & 0.61270 & 0.12227 & 0.00139 & 0.0114 &  \\
OGLE-BLG-RRLYR-07448 & 0.37014 & 0.23241 & 0.62788 & 0.07002 & 0.00177 & 0.0252 & s \\
OGLE-BLG-RRLYR-07486 & 0.31703 & 0.19974 & 0.63005 & 0.14016 & 0.00218 & 0.0156 & a,c,g,s \\
OGLE-BLG-RRLYR-07500 & 0.32308 & 0.20407 & 0.63163 & 0.12655 & 0.00111 & 0.0087 & a,s,c,e \\
& 0.32308 & 0.20081 & 0.62157 & 0.12655 & 0.00128 & 0.0101 &    \\
OGLE-BLG-RRLYR-07517 & 0.32565 & 0.20497 & 0.62943 & 0.12239 & 0.00236 & 0.0193 & a,g,s \\
OGLE-BLG-RRLYR-07518 & 0.29361 & 0.18029 & 0.61405 & 0.13590 & 0.00373 & 0.0275 & a,c,g,h \\
OGLE-BLG-RRLYR-07559 & 0.32052 & 0.19617 & 0.61204 & 0.11084 & 0.00152 & 0.0137 & a,c,g,s \\
& 0.32052 & 0.20239 & 0.63144 & 0.11086 & 0.00124 & 0.0112 &    \\
OGLE-BLG-RRLYR-07677 & 0.29486 & 0.18080 & 0.61318 & 0.11823 & 0.00341 & 0.0288 & g \\
OGLE-BLG-RRLYR-07701 & 0.29796 & 0.18267 & 0.61309 & 0.14771 & 0.00227 & 0.0154 & a,g \\
OGLE-BLG-RRLYR-07714 & 0.36978 & 0.22641 & 0.61227 & 0.12508 & 0.00132 & 0.0106 & a,c,g \\
OGLE-BLG-RRLYR-07723 & 0.35599 & 0.21762 & 0.61131 & 0.13238 & 0.00175 & 0.0132 & a,s \\
OGLE-BLG-RRLYR-07781 & 0.42917 & 0.26464 & 0.61664 & 0.12329 & 0.00194 & 0.0157 & a,td,c \\
OGLE-BLG-RRLYR-07803 & 0.31390 & 0.19220 & 0.61230 & 0.12436 & 0.00285 & 0.0229 & a,c,g \\
OGLE-BLG-RRLYR-07806 & 0.31900 & 0.19820 & 0.62131 & 0.12226 & 0.00190 & 0.0156 & a,c,s,g \\
 & 0.31900 & 0.19547 & 0.61276 & 0.12226 & 0.00168 & 0.0138 &  \\
 & 0.31900 & 0.20132 & 0.63109 & 0.12226 & 0.00149 & 0.0122 &  \\
OGLE-BLG-RRLYR-07857 & 0.32088 & 0.19645 & 0.61224 & 0.12395 & 0.00156 & 0.0126 & a,c,s \\
 & 0.32088 & 0.19938 & 0.62137 & 0.12395 & 0.00222 & 0.0179 &  \\
OGLE-BLG-RRLYR-07907 & 0.28779 & 0.17659 & 0.61360 & 0.14599 & 0.00398 & 0.0273 &  \\
OGLE-BLG-RRLYR-07962 & 0.24463 & 0.14982 & 0.61244 & 0.11238 & 0.00311 & 0.0277 & c,d \\
OGLE-BLG-RRLYR-08002 & 0.30879 & 0.18918 & 0.61264 & 0.11784 & 0.00267 & 0.0227 & a,c,f,g \\
& 0.30879 & 0.19471 & 0.63055 & 0.11784 & 0.00152 & 0.0129 &    \\
OGLE-BLG-RRLYR-08123 & 0.28762 & 0.18132 & 0.63042 & 0.12587 & 0.00150 & 0.0119 & a \\
 & 0.28762 & 0.17583 & 0.61134 & 0.12587 & 0.00199 & 0.0158 &  \\
OGLE-BLG-RRLYR-08125 & 0.27681 & 0.16954 & 0.61246 & 0.10073 & 0.00453 & 0.0450 & a,c,e,g,h \\
 & 0.27681 & 0.17468 & 0.63103 & 0.10073 & 0.00188 & 0.0187 &  \\
OGLE-BLG-RRLYR-08137 & 0.30800 & 0.18914 & 0.61411 & 0.12859 & 0.00296 & 0.0231 & a,c,g \\
OGLE-BLG-RRLYR-08138 & 0.31864 & 0.19525 & 0.61276 & 0.12119 & 0.00131 & 0.0108 & a,e,s,td \\
& 0.31864 & 0.20104 & 0.63093 & 0.12119 & 0.00143 & 0.0118 &    \\
OGLE-BLG-RRLYR-08170 & 0.32306 & 0.19797 & 0.61281 & 0.11671 & 0.00151 & 0.0130 & a,g \\
OGLE-BLG-RRLYR-08177 & 0.28513 & 0.17518 & 0.61438 & 0.14088 & 0.00204 & 0.0145 & bl \\
OGLE-BLG-RRLYR-08302 & 0.29946 & 0.18393 & 0.61421 & 0.13206 & 0.00273 & 0.0207 & a,g \\
OGLE-BLG-RRLYR-08390 & 0.29069 & 0.17800 & 0.61233 & 0.13221 & 0.00192 & 0.0145 & a,c \\
OGLE-BLG-RRLYR-08421 & 0.30188 & 0.18461 & 0.61153 & 0.12594 & 0.00330 & 0.0262 & a,g,h \\
OGLE-BLG-RRLYR-08447 & 0.27066 & 0.16390 & 0.60555 & 0.15470 & 0.00106 & 0.0069 &  \\
OGLE-BLG-RRLYR-08460 & 0.36490 & 0.22846 & 0.62609 & 0.11244 & 0.00177 & 0.0157 & a,c,d,s \\
OGLE-BLG-RRLYR-08590 & 0.24166 & 0.14847 & 0.61438 & 0.08419 & 0.00267 & 0.0317 & a,g \\
OGLE-BLG-RRLYR-08594 & 0.29979 & 0.18356 & 0.61231 & 0.13151 & 0.00519 & 0.0395 & a,c,g \\
OGLE-BLG-RRLYR-08597 & 0.32093 & 0.20228 & 0.63031 & 0.13666 & 0.00279 & 0.0204 & a,s,h \\
OGLE-BLG-RRLYR-08653 & 0.36015 & 0.22232 & 0.61731 & 0.13333 & 0.00131 & 0.0098 & a \\
OGLE-BLG-RRLYR-08674 & 0.31692 & 0.19433 & 0.61319 & 0.12192 & 0.00186 & 0.0152 & a,c,g,s,h \\
OGLE-BLG-RRLYR-08696 & 0.32765 & 0.20421 & 0.62326 & 0.12791 & 0.00125 & 0.0098 & a,td,c,g,s \\
 & 0.32765 & 0.20775 & 0.63407 & 0.12791 & 0.00134 & 0.0104 &  \\
OGLE-BLG-RRLYR-08715 & 0.36986 & 0.23325 & 0.63064 & 0.12879 & 0.00173 & 0.0134 & a \\
OGLE-BLG-RRLYR-08720 & 0.27942 & 0.17212 & 0.61598 & 0.15325 & 0.00135 & 0.0088 & a,c,g,sr \\
OGLE-BLG-RRLYR-08721 & 0.24132 & 0.14756 & 0.61146 & 0.08050 & 0.00410 & 0.0510 & a,c,f,h \\
OGLE-BLG-RRLYR-08824 & 0.30036 & 0.18370 & 0.61162 & 0.12436 & 0.00399 & 0.0321 & a,c,f,g \\
OGLE-BLG-RRLYR-08826 & 0.31251 & 0.19127 & 0.61204 & 0.10170 & 0.00166 & 0.0163 & a,c,s \\
 & 0.31251 & 0.19732 & 0.63140 & 0.10170 & 0.00149 & 0.0146 &  \\
 & 0.31251 & 0.19404 & 0.62089 & 0.10170 & 0.00100 & 0.0099 &  \\
   \hline
 \end{tabular}
 \end{minipage}
 \end{table*}
 
  \begin{table*} 
 \centering
  \begin{minipage}{140mm}
  \begin{tabular}{@{}lccccccl@{}}
   \hline
Name & $\po$\thinspace[d] & $\px$\thinspace[d] & $\px/\po$ & $\ao$\thinspace[mag] & $\ax$\thinspace[mag] & $\ax/\ao$ &   Remarks            \\
  \hline
OGLE-BLG-RRLYR-08844 & 0.28957 & 0.17753 & 0.61307 & 0.12863 & 0.00261 & 0.0203 & g \\
OGLE-BLG-RRLYR-08847 & 0.26403 & 0.16192 & 0.61327 & 0.07226 & 0.00172 & 0.0238 & g \\
OGLE-BLG-RRLYR-08863 & 0.28168 & 0.17291 & 0.61388 & 0.14696 & 0.00175 & 0.0119 & a,s \\
OGLE-BLG-RRLYR-08866 & 0.24142 & 0.14800 & 0.61304 & 0.13513 & 0.00227 & 0.0168 & g \\
OGLE-BLG-RRLYR-08986 & 0.30559 & 0.18777 & 0.61445 & 0.12067 & 0.00303 & 0.0251 &  \\
OGLE-BLG-RRLYR-09126 & 0.32399 & 0.20138 & 0.62156 & 0.12484 & 0.00282 & 0.0226 & a,c,g,s \\
 & 0.32399 & 0.19847 & 0.61260 & 0.12484 & 0.00231 & 0.0185 &  \\
 & 0.32399 & 0.20468 & 0.63177 & 0.12484 & 0.00143 & 0.0115 &  \\
OGLE-BLG-RRLYR-09164 & 0.28639 & 0.17558 & 0.61309 & 0.14306 & 0.00307 & 0.0214 & c,g,h \\
OGLE-BLG-RRLYR-09212 & 0.23495 & 0.14436 & 0.61441 & 0.12440 & 0.00178 & 0.0143 & a,g \\
OGLE-BLG-RRLYR-09305 & 0.30295 & 0.18586 & 0.61350 & 0.12718 & 0.00428 & 0.0337 & a,c,g,s,h \\
 & 0.30295 & 0.18858 & 0.62248 & 0.12718 & 0.00153 & 0.0121 &  \\
 & 0.30295 & 0.19126 & 0.63132 & 0.12718 & 0.00171 & 0.0135 &  \\
OGLE-BLG-RRLYR-09436 & 0.33013 & 0.20362 & 0.61678 & 0.12507 & 0.00178 & 0.0142 & a,c,g \\
OGLE-BLG-RRLYR-09511 & 0.30283 & 0.18971 & 0.62648 & 0.11040 & 0.00161 & 0.0146 & a \\
OGLE-BLG-RRLYR-09520 & 0.24893 & 0.15259 & 0.61300 & 0.10905 & 0.00154 & 0.0142 & e \\
OGLE-BLG-RRLYR-09521 & 0.32318 & 0.19808 & 0.61291 & 0.11586 & 0.00133 & 0.0115 & a,s \\
 & 0.32318 & 0.20428 & 0.63209 & 0.11586 & 0.00114 & 0.0098 &  \\
OGLE-BLG-RRLYR-09529 & 0.30693 & 0.18842 & 0.61390 & 0.12524 & 0.00383 & 0.0306 & a,c,d,f,h \\
OGLE-BLG-RRLYR-09631 & 0.28027 & 0.17236 & 0.61498 & 0.14639 & 0.00127 & 0.0087 & a,g \\
OGLE-BLG-RRLYR-09649 & 0.30893 & 0.19014 & 0.61547 & 0.11701 & 0.00376 & 0.0321 & a,g \\
OGLE-BLG-RRLYR-09665 & 0.29925 & 0.18352 & 0.61325 & 0.11705 & 0.00381 & 0.0325 & g \\
OGLE-BLG-RRLYR-09696 & 0.30839 & 0.18772 & 0.60872 & 0.11066 & 0.00069 & 0.0063 & a,g \\
OGLE-BLG-RRLYR-09775 & 0.23541 & 0.14451 & 0.61389 & 0.09353 & 0.00174 & 0.0186 &  \\
OGLE-BLG-RRLYR-09795 & 0.31620 & 0.19441 & 0.61485 & 0.15930 & 0.00127 & 0.0080 & a,g,c \\
OGLE-BLG-RRLYR-10000 & 0.29739 & 0.18269 & 0.61431 & 0.12806 & 0.00246 & 0.0192 & a,d \\
OGLE-BLG-RRLYR-10008 & 0.30435 & 0.18729 & 0.61536 & 0.12697 & 0.00259 & 0.0204 & a,f \\
OGLE-BLG-RRLYR-10037 & 0.32980 & 0.20776 & 0.62997 & 0.13112 & 0.00278 & 0.0212 & a,c,g,s,h \\
OGLE-BLG-RRLYR-10040 & 0.31351 & 0.19233 & 0.61345 & 0.13149 & 0.00200 & 0.0152 & a,c,d,f \\
OGLE-BLG-RRLYR-10127 & 0.28186 & 0.17317 & 0.61438 & 0.14131 & 0.00219 & 0.0155 & a \\
OGLE-BLG-RRLYR-10371 & 0.28935 & 0.17718 & 0.61235 & 0.13613 & 0.00243 & 0.0178 & a,c,g,e,s,h \\
OGLE-BLG-RRLYR-30633 & 0.29150 & 0.17891 & 0.61376 & 0.12989 & 0.00287 & 0.0221 &  \\
OGLE-BLG-RRLYR-30848 & 0.30693 & 0.18930 & 0.61676 & 0.12281 & 0.00234 & 0.0190 & a,g \\
OGLE-BLG-RRLYR-31736 & 0.30524 & 0.18726 & 0.61348 & 0.13337 & 0.00207 & 0.0155 & a,c,g,s \\
 & 0.30524 & 0.19250 & 0.63065 & 0.13337 & 0.00157 & 0.0118 &  \\
OGLE-BLG-RRLYR-32091 & 0.28894 & 0.17691 & 0.61225 & 0.12991 & 0.00353 & 0.0271 & a,t \\
OGLE-BLG-RRLYR-32145 & 0.30005 & 0.18422 & 0.61397 & 0.11418 & 0.00340 & 0.0298 & a,c,g \\
OGLE-BLG-RRLYR-32252 & 0.36310 & 0.22868 & 0.62980 & 0.11697 & 0.00207 & 0.0177 & bl,c \\
OGLE-BLG-RRLYR-32289 & 0.30627 & 0.18845 & 0.61529 & 0.12610 & 0.00265 & 0.0210 & a,c,g \\
OGLE-BLG-RRLYR-32877 & 0.25066 & 0.15337 & 0.61186 & 0.07322 & 0.00349 & 0.0477 &  \\
 & 0.25066 & 0.15600 & 0.62235 & 0.07322 & 0.00247 & 0.0337 &  \\
   \hline
 \end{tabular}
 \end{minipage}
 \end{table*}

\begin{table*}
\centering
\caption{Light curve solution for OGLE-BLG-RRLYR-08177.}
\label{tab:bl1}
\begin{tabular}{lrrrrr}
\hline 
freq. id & $f[{\rm d}^-1]$ & $A$[mag] & $\sigma$ & $\phi$[rad] & $\sigma$ \\
\hline     
$f_{\rm BL}$       & 0.02202965 & 0.0017 & 0.0003 & 0.12 & 0.17 \\
$\fo-f_{\rm BL}$  &  3.48509951 & 0.0334 & 0.0003 & 3.559 & 0.083 \\
$\fo$           &  3.50712916 & 0.1408 & 0.0003 & 2.889 & 0.023 \\
$\fo+f_{\rm BL}$  &  3.52915881 & 0.0256 & 0.0003 & 6.215 & 0.090 \\
$\fx$            &  5.70840168 & 0.0020 & 0.0003 & 3.0 & 1.9 \\
$2\fo-2f_{\rm BL}$ & 6.97019902 & 0.0017 & 0.0003 & 3.71 & 0.22 \\
$2\fo-f_{\rm BL}$ &  6.99222868 & 0.0127 & 0.0003 & 2.165 & 0.091 \\
$2\fo$          &  7.01425833 & 0.0200 & 0.0003 & 2.665 & 0.048 \\
$2\fo+f_{\rm BL}$ &  7.03628798 & 0.0059 & 0.0003 & 5.64 & 0.11 \\
$2\fo+2f_{\rm BL}$ & 7.05831763 & 0.0014 & 0.0003 & 2.83 & 0.26 \\
$3\fo-2f_{\rm BL}$& 10.47732819 & 0.0028 & 0.0003 & 1.53 & 0.19 \\
$3\fo-f_{\rm BL}$ & 10.49935784 & 0.0057 & 0.0003 & 1.90 & 0.11 \\
$3\fo$          & 10.52138749 & 0.0140 & 0.0003 & 2.632 & 0.071 \\
$3\fo+f_{\rm BL}$ & 10.54341714 & 0.0029 & 0.0003 & 6.04 & 0.15 \\
$4\fo-f_{\rm BL}$ & 14.00648700 & 0.0031 & 0.0003 & 1.93 & 0.14 \\
$4\fo$          & 14.02851665 & 0.0085 & 0.0003 & 2.676 & 0.096 \\
$4\fo+f_{\rm BL}$ & 14.05054631 & 0.0028 & 0.0003 & 6.11 & 0.16 \\
$5\fo-f_{\rm BL}$ & 17.51361617 & 0.0019 & 0.0003 & 1.77 & 0.19 \\
$5\fo$          & 17.53564582 & 0.0050 & 0.0003 & 2.22 & 0.12 \\
$5\fo+f_{\rm BL}$ & 17.55767547 & 0.0023 & 0.0003 & 5.55 & 0.19 \\
$6\fo$          & 21.04277498 & 0.0030 & 0.0003 & 1.52 & 0.16 \\
$7\fo$          & 24.54990414 & 0.0015 & 0.0003 & 0.64 & 0.23 \\
\hline 
\hline 
\end{tabular}
\end{table*}

\begin{table*}
\centering
\caption{Light curve solution for OGLE-BLG-RRLYR-32252.}
\label{tab:bl2}
\begin{tabular}{lrrrrr}
\hline 
freq. id & $f[{\rm d}^-1]$ & $A$[mag] & $\sigma$ & $\phi$[rad] & $\sigma$ \\
\hline     
$\fo-f_{\rm BL}$  &  2.75202958 & 0.0073 & 0.0003 & 4.69 & 0.37 \\
$\fo$           &  2.75405244 & 0.1169 & 0.0002 & 4.806 & 0.048 \\
$\fo+f_{\rm BL}$  &  2.75607531 & 0.0080 & 0.0003 & 1.61 & 0.43 \\
$\fx$           &  4.37567732 & 0.0020 & 0.0002 & 0.7 & 1.3 \\
$2\fo$          &  5.50810489 & 0.0067 & 0.0002 & 1.01 & 0.10 \\
$\fo+\fx$       &  7.12972977 & 0.0017 & 0.0002 & 0.6 & 1.3 \\
$3\fo-f_{\rm BL}$ &  8.26013447 & 0.0017 & 0.0002 & 3.16 & 0.35 \\
$3\fo$          &  8.26215733 & 0.0075 & 0.0002 & 3.50 & 0.15 \\
$3\fo+f_{\rm BL}$ &  8.26418020 & 0.0018 & 0.0002 & 0.16 & 0.51 \\
$4\fo$          & 11.01620978 & 0.0043 & 0.0002 & 5.42 & 0.20 \\
$4\fo+f_{\rm BL}$ & 11.01823264 & 0.0012 & 0.0002 & 2.17 & 0.57 \\
$5\fo$          & 13.77026222 & 0.0025 & 0.0002 & 0.91 & 0.26 \\
$6\fo$          & 16.52431467 & 0.0013 & 0.0002 & 2.54 & 0.34 \\

\hline 
\hline 
\end{tabular}
\end{table*}


\bsp	
\label{lastpage}
\end{document}